\documentclass[11pt,leqno,twoside]{article}

\usepackage[paperwidth=8.5in,paperheight=11in,textwidth=405bp,textheight=646bp,centering]{geometry}
\usepackage{amsmath,amssymb,amsthm,mathtools}
\usepackage{bm}
\usepackage{booktabs}
\usepackage{array}
\usepackage{enumitem}
\usepackage{natbib}
\usepackage{graphicx}
\usepackage{longtable}
\usepackage{float}
\usepackage{placeins}
\usepackage{fullpage}
\usepackage{hyperref}
\hypersetup{
  colorlinks=true,
  linkcolor=blue,
  citecolor=blue,
  urlcolor=blue
}

\allowdisplaybreaks

\numberwithin{equation}{section}

\newtheorem{theorem}{Theorem}[section]
\newtheorem{proposition}{Proposition}[section]

\theoremstyle{definition}
\newtheorem{assumption}{Assumption}[section]
\theoremstyle{remark}
\newtheorem{remark}{Remark}[section]

\newcommand{\E}{\mathbb E}
\newcommand{\Pp}{\mathbb P}

\newcommand{\N}{\mathbb N}
\newcommand{\Var}{\operatorname{var}}
\newcommand{\Cov}{\operatorname{cov}}
\newcommand{\cum}{\operatorname{cum}}
\newcommand{\sgn}{\operatorname{sign}}

\newcommand{\ind}{\mathbf 1}
\newcommand{\Ho}{H_0}

\newcommand{\Lam}{\Lambda_p}
\newcommand{\npairs}{N_p}

\newcommand{\wt}{\widetilde}
\newcommand{\calQ}{\mathcal Q}
\newcommand{\calB}{\mathcal B}
\newcommand{\calJ}{\mathcal J}

\newcommand{\calL}{\mathcal L}

\newcommand{\iid}{\stackrel{\mathrm{iid}}{\sim}}

\newcommand{\Sf}{\mathfrak S}
\newcommand{\fall}[2]{(#1)_{#2}}
\newcommand{\vct}[1]{\bm{#1}}

\begin{document}

\begin{center}
{\Large\bf Rank-Based Tests for Mutual Independence of High-Dimensional Random Vectors via $L_q$ Norm}\par
\vspace{0.5cm}
{\sc Ping Zhao, Hongfei Wang and Long Feng}\par
\vspace{0.2cm}
{\it Tiangong University, Nanjing Audit University, Nankai University}\par
\vspace{0.25cm}
\end{center}

\begin{abstract}
We consider the problem of testing mutual independence among the components of a high-dimensional random vector.  Building on the rank-based max-sum framework, we introduce fixed finite-$L_q$ power-sum statistics under three general classes of rank-based correlations: simple linear rank statistics, non-degenerate rank-based U-statistics and degenerate rank-based U-statistics.  The proposed statistics interpolate between the dense-alternative sensitivity of the $L_2$ statistic and the sparse-alternative sensitivity of the $L_\infty$ statistic.  We establish the asymptotic independence between any fixed finite-$L_q$ block and the corresponding $L_\infty$ statistic, and combine $L_2,L_4,L_6$ and $L_\infty$ p-values through a Cauchy rule.  Numerical studies show that the resulting $L_{2,4,6,\infty}$ procedure is highly robust to the sparsity of the alternative and has strong empirical power across the considered designs.
\end{abstract}

\noindent{\bf Keywords:} asymptotic independence; Cauchy combination; finite-$L_q$ tests; high-dimensional independence; rank correlation; U-statistics.

\medskip
\noindent{\bf MSC2020 subject classifications:} Primary 62G10, 62G20; secondary 62H15, 62H20.

\section{Introduction}

Let
\[
\bm{X}=(X_1,\ldots,X_p)^\top
\]
be a $p$-dimensional random vector, and suppose that independent copies
\[
\bm{X}_i=(X_{i1},\ldots,X_{ip})^\top,\qquad i=1,\ldots,n,
\]
are observed.  We study the global mutual independence hypothesis
\begin{equation}\label{eq:H0}
\Ho: X_1,\ldots,X_p \hbox{ are mutually independent}.
\end{equation}
The dimension $p$ may be comparable to, or substantially larger than, the sample size $n$.  This problem is central in high-dimensional inference because it determines whether a large collection of variables can be analyzed as mutually unrelated.  It arises in network screening, genomics, neuroscience, finance and econometrics, where falsely imposing independence can distort uncertainty quantification, edge selection and risk assessment.

In low-dimensional Gaussian models, classical likelihood-ratio and covariance-matrix procedures are available; see, for example, \citet{Anderson2003}, \citet{Roy1954} and \citet{Nagao1973}.  These procedures are not directly applicable when $p>n$, because the sample covariance matrix is singular.  To address high-dimensional covariance and correlation testing, \citet{BaiJiangYaoZheng2009}, \citet{JiangJiangYang2012} and \citet{JiangYang2013} proposed likelihood-ratio adaptations based on random matrix theory.  Sum-type procedures aggregate squared sample correlations and are naturally sensitive to dense alternatives; \citet{Schott2005} proposed a modified Nagao-type statistic based on Pearson correlations.  Maximum-type procedures use the largest sample correlation and are designed for sparse alternatives; \citet{Jiang2004}, \citet{LiuLinShao2008} and \citet{CaiJiang2011} studied the corresponding extreme-value approximations for high-dimensional correlation matrices.  Because the sparsity level of the unknown dependence structure is rarely known in applications, \citet{LiXue2015} and \citet{FengJiangLiuXiong2022} proposed adaptive max-sum procedures that combine sum-type and maximum-type statistics.

Pearson-correlation procedures can be sensitive to marginal distributions, heavy tails and nonlinear dependence.  A substantial body of work therefore replaces Pearson correlation by distribution-free or nonlinear dependence coefficients.  \citet{WangZouWang2013} proposed a necessary rank-correlation test for complete independence in high dimensions.  \citet{Mao2014} and \citet{Mao2017} developed nonparametric and robust high-dimensional independence tests, and \citet{Mao2018} studied a Kendall's $\tau$ based complete-independence test.  \citet{ShiXuDu2023} proposed quadratic and maximum statistics based on Spearman's footrule and combined them into a max-sum test.  In the three-class rank framework used in this paper, classical examples include Spearman's $\rho$ \citep{Spearman1904}, Kendall's $\tau$ \citep{Kendall1938}, Hoeffding's $D$ \citep{Hoeffding1948b}, Blum--Kiefer--Rosenblatt's $R$ \citep{BlumKieferRosenblatt1961} and Bergsma--Dassios--Yanagimoto's $\tau^\ast$ \citep{BergsmaDassios2014}.  \citet{HanChenLiu2017} derived maximum-type rank tests for simple linear rank statistics and non-degenerate rank-based U-statistics.  \citet{LeungDrton2018} derived sum-type rank tests for non-degenerate and degenerate rank-based U-statistics.  \citet{DrtonHanShi2020} established maximum-type theory for degenerate rank correlations.  \citet{WangLiuFengMa2024} unified these three classes through rank-based max-sum tests and proved the asymptotic independence between the rank-based $L_2$ sum statistic and the $L_\infty$ maximum statistic.

Other nonparametric coefficients provide complementary high-dimensional independence tools.  \citet{SzekelyRizzoBakirov2007} introduced distance covariance and distance correlation, which characterize general dependence between random vectors.  \citet{SzekelyRizzo2013} developed a high-dimensional distance-correlation $t$-test for independence between random vectors, and \citet{YaoZhangShao2018} constructed an $L_2$-type high-dimensional mutual-independence test based on pairwise distance covariance.  Kernel-based approaches are also closely related: \citet{GrettonEtAl2005} introduced Hilbert--Schmidt norm measures of dependence, \citet{GrettonEtAl2008} proposed the HSIC independence test, and \citet{PfisterEtAl2018} extended HSIC to joint independence through dHSIC.  \citet{HellerHellerGorfine2013} developed a multivariate distance-rank test of association, and \citet{PanEtAl2020} introduced ball covariance as a generic metric-space dependence measure.  More recently, \citet{Chatterjee2021} proposed a rank correlation coefficient that is consistent against all alternatives, \citet{XiaCaoDuDai2025} developed quadratic and extreme-value complete-independence tests based on Chatterjee's coefficient, and \citet{XiaCaoDuLiu2025} combined Spearman's $\rho$ and Chatterjee's $\xi$ for high-dimensional independence testing.

Finite $L_q$ norms offer a systematic way to interpolate between dense and sparse alternatives.  The $L_2$ statistic is most effective when many coordinates of the dependence structure are nonzero, whereas the $L_\infty$ statistic is most effective when the signal is concentrated in a small number of coordinates.  Fixed finite orders such as $q=4$ and $q=6$ target moderately sparse alternatives.  \citet{XuLinWeiPan2016} and \citet{WuXuPan2019} studied sums of powers and related $L_q$-type procedures for high-dimensional testing.  \citet{HeEtAl2021} constructed unbiased U-statistics for powers of high-dimensional parameters and established asymptotic independence across different $q$'s and with maximum-type statistics in several settings.  \citet{ZhangWangShao2025} developed a general $L_q$-norm based U-statistic theory for kernels of order larger than one, including component-wise independence examples based on Kendall's $\tau$ and Spearman's $\rho$.  Their results give asymptotic normality and joint independence for several fixed $q$'s and show that larger finite $q$ values are advantageous as the alternative becomes sparser.

This paper extends the rank-based max-sum framework from the two endpoint statistics, $L_2$ and $L_\infty$, to a finite-$L_q$ spectrum.  We work under the same three rank-based classes as \citet{WangLiuFengMa2024}; the named coefficients above serve only as representative examples.  For U-statistic-based members of the framework, we use the finite-$L_q$ Gaussian approximation of \citet{ZhangWangShao2025}.  The new theoretical component is a rank-specific block independence result: under the null hypothesis, any fixed finite collection of rank-based $L_q$ statistics is asymptotically independent of the corresponding $L_\infty$ statistic.  Building on \citet{LiuXie2020}, we then construct a Cauchy combination test that aggregates $L_2,L_4,L_6$ and $L_\infty$ p-values.

The main contributions are as follows.  We formulate fixed finite-$L_q$ power-sum tests under the three general classes of rank-based correlations.  We prove block asymptotic independence between the finite-$L_q$ vector and the maximum statistic, with explicit rate bounds that display the required relationship between $p$ and $n$.  We construct an adaptive Cauchy $L_{2,4,6,\infty}$ procedure that retains sensitivity to dense, moderately sparse and very sparse alternatives.  For implementation, we give closed finite-sample centering and variance formulas for Spearman's $\rho$ and Kendall's $\tau$, and use high-precision null simulation for the degenerate rank correlations, whose high-order finite-sample moments are combinatorially much more complex.

The remainder of the paper is organized as follows.  Section \ref{sec:method} presents the methodology and main theory, including the three rank-based classes, finite-$L_q$ statistics, assumptions, null calibration, block independence, Cauchy combination and representative calibration formulas.  Section \ref{sec:simulation} describes the simulation design for assessing empirical size and power across sparsity regimes.  Section \ref{sec:discussion} concludes.  Proofs are collected in the Appendix.

\paragraph{Notation.} In this paper, we represent random vectors by uppercase letters (e.g., $X_i$) and their individual components by the corresponding lowercase letters (e.g., $x_{ij}$). For any positive integer $n$, we let $[n]$ denote the set $\{1, \ldots, n\}$, and $[m, n]$ represent the range $\{m, m+1, \ldots, n\}$. Given a sequence of (not necessarily distinct) indices $I = (i_1, i_2, \ldots, i_m)$ from $[n]$, we write $X_I$ for the concatenated vector $(X_{i_1}, X_{i_2}, \ldots, X_{i_m})$; in particular, $X_{[c]}$ is defined as $(X_1, \ldots, X_c)$. We employ $\sum^*$ to signify summation restricted to mutually distinct indices. For functions $f(X_1, \ldots, X_r)$ and $g(Y_1, \ldots, Y_s)$, their tensor product $f \otimes g$ is defined as the pointwise product $f(X_1, \ldots, X_r)g(Y_1, \ldots, Y_s)$, with the recursive form $(\otimes)^s f = f \otimes [(\otimes)^{s-1} f]$ for $s \geq 2$. The collection of the distinct $m$-tuples from $[n]$ is denoted by $P_m(n)=\{(i_1,\ldots,i_m)\in[n]^m:i_1,\ldots,i_m\text{ are pairwise distinct}\}$. Asymptotic equivalence $f \asymp g$ is used to indicate that $f(n)/g(n) \to 1$ as $n \to \infty$. Regarding vector operations in $\mathbb{R}^p$, we define the $q$-th power sum as $\|X\|_q^q = \sum_{l=1}^p x_l^q$ for $q \in \mathbb{Z}_{+}$, which differs from the standard $L_q$ norm for odd $q$ as the absolute value is omitted. The standard $L_\infty$ norm is denoted by $\|X\|_{\infty} = \max_{l=1}^p |x_l|$, and we simplify $\| \cdot \|_2$ to $\| \cdot \|$. Finally, we reserve $e_t(r)$ for the $t$-th standard basis vector in $\mathbb{R}^r$ and $\boldsymbol{I}_k$ for the $k \times k$ identity matrix.  Let $\Sf_m$ be the permutation group on $m$ elements.
\section{Methodology and main theory}\label{sec:method}

\subsection{Three rank-based classes and finite-$L_q$ statistics}\label{sec:setup}

Let
\[
\Lam=\{(s,t):1\le s<t\le p\},\qquad \npairs=|\Lam|={p\choose2}.
\]
All expectations, variances and probabilities with subscript $0$ are taken under \eqref{eq:H0}.  We assume throughout that the marginal distributions are continuous, so ranks are almost surely free of ties.

\subsubsection{Simple linear rank statistics}\label{subsec:linear}

Let $R^s_{ni}$ be the rank of $X_{is}$ among $X_{1s},\ldots,X_{ns}$.  For score functions $f,g:(0,1)\to\mathbb R$, define
\begin{equation}\label{eq:linear_rank}
V_{st}=\frac1n\sum_{i=1}^n f\left(\frac{R^s_{ni}}{n+1}\right)g\left(\frac{R^t_{ni}}{n+1}\right),\qquad (s,t)\in\Lam.
\end{equation}
The statistic $V_{st}$ measures agreement between the two rank sequences $(R^s_{n1},\ldots,R^s_{nn})$ and $(R^t_{n1},\ldots,R^t_{nn})$ through the scores $f$ and $g$.  Under $H_0$, the two rank vectors are independent uniform permutations of $1,\ldots,n$.  Without loss of generality one may replace $f$ and $g$ by their centered discrete versions
\[
f_{ni}=f\left(\frac{i}{n+1}\right)-\bar f_n,
\qquad
 g_{ni}=g\left(\frac{i}{n+1}\right)-\bar g_n,
\]
where
\[
\bar f_n=\frac1n\sum_{i=1}^n f\left(\frac{i}{n+1}\right),
\qquad
\bar g_n=\frac1n\sum_{i=1}^n g\left(\frac{i}{n+1}\right).
\]
This class includes Spearman's $\rho$ \citep{Spearman1904} after a deterministic normalization of the linear scores.

\subsubsection{Rank-based U-statistics and degeneracy}\label{subsec:u_general}

Let $h:\mathbb R^{2m}\to\mathbb R$ be a bounded, symmetric, rank-based kernel.  For $\bm{X}_i^{(st)}=(X_{is},X_{it})^\top$, define
\begin{equation*}
U_{h,st}={n\choose m}^{-1}\sum_{1\le i_1<\cdots<i_m\le n}
h\{\bm{X}_{i_1}^{(st)},\ldots,\bm{X}_{i_m}^{(st)}\}.
\end{equation*}
Rank-based means that $h$ depends on its arguments only through their coordinate-wise ranks among the $m$ points.  If an original kernel is not symmetric, it may be replaced by its symmetrization without changing the associated U-statistic.

For $1\le r\le m$, define the $r$th projection
\begin{equation*}
h_r(z_1,\ldots,z_r)=\E_0\{h(z_1,\ldots,z_r,Z_{r+1},\ldots,Z_m)\},
\end{equation*}
where $\bm{Z}_1,\ldots,\bm{Z}_m$ are iid copies of $\bm{X}^{(st)}=(X_s,X_t)^\top$ under $H_0$.  The kernel has degeneracy order $d\in\{1,\ldots,m\}$ if
\begin{equation*}
h_1=\cdots=h_{d-1}=0,
\qquad
h_d\not\equiv 0.
\end{equation*}
The case $d=1$ is non-degenerate.  The case $d\ge2$ is degenerate.  The present paper follows \citet{WangLiuFengMa2024} and focuses on the non-degenerate class $d=1$ and the degenerate class $d=2$.

\subsubsection{Non-degenerate rank-based U-statistics}\label{subsec:nondeg}

When $d=1$, write
\begin{equation}\label{eq:nondeg_u}
W_{h,st}=U_{h,st},
\qquad (s,t)\in\Lam.
\end{equation}
The first projection controls both the finite-$L_q$ sum statistic and the maximum statistic.  The factorization condition used in rank-based maximum theory requires
\begin{equation*}
h_1(x_s,x_t)=\omega(x_s)\gamma(x_t)
\end{equation*}
for measurable functions $\omega$ and $\gamma$.  Kendall's $\tau$ \citep{Kendall1938} is the canonical example with $m=2$ and
\[
h_\tau\{\bm{X}_{i_1}^{(st)},\bm{X}_{i_2}^{(st)}\}=\sgn\{(X_{i_1s}-X_{i_2s})(X_{i_1t}-X_{i_2t})\}.
\]

\subsubsection{Degenerate rank-based U-statistics}\label{subsec:deg}

When $d=2$, write
\begin{equation}\label{eq:deg_u}
Q_{h,st}=U_{h,st},
\qquad (s,t)\in\Lam.
\end{equation}
Let $h_2$ be the second-order projection.  The degenerate maximum theory assumes the spectral representation
\begin{equation*}
h_2(x,y)=\sum_{v=1}^{\infty}\lambda_v\phi_v(x)\phi_v(y),
\qquad \lambda_1\ge\lambda_2\ge\cdots\ge0,
\end{equation*}
where
\begin{equation*}
\Lambda=\sum_{v=1}^{\infty}\lambda_v<\infty,
\qquad
\sup_{v\ge1}\|\phi_v\|_\infty<\infty.
\end{equation*}
Let $\mu_1$ be the multiplicity of $\lambda_1$, and define
\begin{equation}\label{eq:kappa_general}
\kappa=\prod_{v>\mu_1}\left(1-\frac{\lambda_v}{\lambda_1}\right)^{-1/2}.
\end{equation}
Hoeffding's $D$ \citep{Hoeffding1948b}, Blum--Kiefer--Rosenblatt's $R$ \citep{BlumKieferRosenblatt1961}, and Bergsma--Dassios--Yanagimoto's $\tau^\ast$ \citep{BergsmaDassios2014} are examples of this class.

\subsubsection{Finite-$L_q$ power sums}\label{subsec:lq_defs}

Let $A_{st}$ denote a generic member of one of the three classes:
\[
A_{st}\in\{V_{st},W_{h,st},Q_{h,st}\}.
\]
Write
\[
\wt A_{st}=A_{st}-\E_0(A_{st}).
\]
For a fixed even integer $q\ge2$, define the single-pair centering and variance
\begin{equation*}
\mu_{A,q,n}=\E_0(\wt A_{12}^{\,q}),
\qquad
v_{A,q,n}=\Var_0(\wt A_{12}^{\,q}),
\end{equation*}
and the global finite-$L_q$ statistic
\begin{equation}\label{eq:lq_stat}
S_{A,q}=\sum_{(s,t)\in\Lam}\{\wt A_{st}^{\,q}-\mu_{A,q,n}\},
\qquad
Z_{A,q}=\frac{S_{A,q}}{(\npairs v_{A,q,n})^{1/2}}.
\end{equation}
The statistic $Z_{A,2}$ is the usual sum statistic.  The values $q=4$ and $q=6$ are intermediate finite-$L_q$ statistics between $L_2$ and $L_\infty$.

\begin{remark}\label{rem:powersum}
For the U-statistic classes, $\wt A_{st}^{\,q}$ is a V-statistic of order $qm$.  Its diagonal-free version is the $L_q$ U-statistic studied by \citet{ZhangWangShao2025}.  Under bounded kernels and fixed $q$, the diagonal-deletion difference is of smaller order after the normalization in \eqref{eq:lq_stat}.  We keep the power-sum form because it is the direct finite-$L_q$ analogue of the $L_2$ statistic in the rank-based max-sum paper and because it yields transparent centering and scaling formulas for concrete rank coefficients.
\end{remark}

\subsection{Mathematical assumptions}\label{sec:assumptions}

We state the assumptions in a form parallel to the rank-based max-sum paper and the general $L_q$ U-statistic theory.

\begin{assumption}\label{ass:A0}
As $n,p\to\infty$,
\begin{equation*}
X_1,\ldots,X_p \hbox{ have continuous marginal distributions},
\end{equation*}

\begin{equation*}
\npairs={p\choose2}\to\infty,
\qquad
\calQ=\{q_1,\ldots,q_K\}\subset2\mathbb{Z}^{+}
\quad\hbox{is fixed}.
\end{equation*}
\end{assumption}

\begin{assumption}\label{ass:V}
For $V_{st}$ in \eqref{eq:linear_rank}, there exist constants $C,C_1,C_2<\infty$ such that
\begin{equation*}
\|f\|_\infty+\|g\|_\infty\le C,
\qquad
\sum_{i=1}^{n}\left[g\left(\frac{i}{n+1}\right)-\bar g_n\right]^2>0,
\end{equation*}
\begin{equation*}
\log p=o(n^{1/3}),
\end{equation*}
\begin{equation*}
f \hbox{ is differentiable and } \sup_{x\in(0,1)}|f'(x)|\le C,
\end{equation*}
\begin{equation*}
\max_{1\le i\le n}\left|g\left(\frac{i}{n+1}\right)-\bar g_n\right|^2
\le
\frac{C_1^2}{n}\sum_{i=1}^{n}\left[g\left(\frac{i}{n+1}\right)-\bar g_n\right]^2,
\end{equation*}
\begin{equation*}
\left|\sum_{i=1}^{n}\left[g\left(\frac{i}{n+1}\right)-\bar g_n\right]^3\right|^2
\le
\frac{C_2^2}{n}
\left\{\sum_{i=1}^{n}\left[g\left(\frac{i}{n+1}\right)-\bar g_n\right]^2\right\}^3.
\end{equation*}
\end{assumption}

\begin{assumption}\label{ass:W}
For $W_{h,st}$ in \eqref{eq:nondeg_u},
\begin{equation*}
\|h\|_\infty<\infty,
\qquad
h \hbox{ is symmetric and rank-based},
\qquad
\E_0\{h(\bm{X}_1^{(st)},\ldots,\bm{X}_m^{(st)})\}=0,
\end{equation*}
\begin{equation*}
\zeta_1^h:=\Var_0\{h_1(\bm{X}_1^{(st)})\}>0,
\end{equation*}
\begin{equation*}
\E_0\{h(\bm{X}_i^{(st)})\mid \bm{X}_j^{(s)},\bm{X}_{j'}^{(t)}\}=0
\quad\hbox{whenever } j,j'\subset i,
\quad \min(|j|,|j'|)<1,
\end{equation*}
\begin{equation*}
h_1(x_s,x_t)=\omega(x_s)\gamma(x_t),
\qquad
\E_0\{\omega^2(X_s)\}\E_0\{\gamma^2(X_t)\}>0,
\end{equation*}
\begin{equation*}
\log p=o(n^{1/3}).
\end{equation*}
\end{assumption}

\begin{assumption}\label{ass:Q}
For $Q_{h,st}$ in \eqref{eq:deg_u},
\begin{equation*}
\|h\|_\infty<\infty,
\qquad
h \hbox{ is symmetric and rank-based},
\qquad
\E_0\{h(\bm{X}_1^{(st)},\ldots,\bm{X}_m^{(st)})\}=0,
\end{equation*}
\begin{equation*}
h_1\equiv0,
\qquad
\zeta_2^h:=\Var_0\{h_2(\bm{X}_1^{(st)},\bm{X}_2^{(st)})\}>0,
\end{equation*}
\begin{equation*}
\E_0\{h(\bm{X}_i^{(st)})\mid \bm{X}_j^{(s)},\bm{X}_{j'}^{(t)}\}=0
\quad\hbox{whenever } j,j'\subset i,
\quad \min(|j|,|j'|)<2,
\end{equation*}
\begin{equation*}
 h_2(x,y)=\sum_{v\ge1}\lambda_v\phi_v(x)\phi_v(y),
 \quad \lambda_1\ge\lambda_2\ge\cdots\ge0,
 \quad \sum_{v\ge1}\lambda_v<\infty,
 \quad \sup_v\|\phi_v\|_\infty<\infty,
\end{equation*}
\begin{equation*}
\phi_v(x_s,x_t)=\psi_v(x_s)\varphi_v(x_t),
\qquad v\ge1.
\end{equation*}
If infinitely many $\lambda_v$ are nonzero, assume
\begin{equation}\label{eq:theta_def}
\log p=o(n^\theta),
\qquad
\theta<\sup\left\{k\in[0,1/3):
\sum_{v>\lfloor n^{(1-3k)/5}\rfloor}\lambda_v=O(n^{-k})\right\}.
\end{equation}
If only finitely many $\lambda_v$ are nonzero, assume \eqref{eq:theta_def} with any $\theta<1/3$.
\end{assumption}

\begin{assumption}\label{ass:LqU}
For a U-statistic class with kernels $h_\ell$, $\ell\in\calL=\Lam$, degeneracy order $s\in\{1,2\}$, and every fixed $q\in\calQ$, let
\[
\sigma_s(\ell_1,\ell_2)=\Cov_0\{h^{(s)}_{\ell_1}(X_{[s]}),h^{(s)}_{\ell_2}(X_{[s]})\},
\]
\[
\Delta_s(q)=\sum_{\ell_1,\ell_2\in\calL}\sigma_s(\ell_1,\ell_2)^q,
\qquad
\widetilde\Delta_s(q)=\Var_0\left\{\sum_{\ell\in\calL}H^{(s)}_{\ell,q}(X_{[qs]})\right\},
\]
where $h^{(s)}_\ell$ is the $s$th Hoeffding projection and $H^{(s)}_{\ell,q}$ is the symmetrization of $\otimes^q h^{(s)}_\ell$ as in \citet{ZhangWangShao2025}.  The following conditions hold:
\begin{equation}\label{eq:Lq_leading}
\Delta_t(q)=o\{n^{q(t-s)}\Delta_s(q)\},\quad s<t\le m,
\qquad
\Delta_s(q)=O\{\widetilde\Delta_s(q)\},
\end{equation}
\begin{equation*}
\sum_{\ell_1,\ldots,\ell_4\in\calL}
\left|\sigma_s(\ell_1,\ell_2)\sigma_s(\ell_3,\ell_4)
\sigma_s(\ell_1,\ell_4)\sigma_s(\ell_2,\ell_3)\right|^{q/2}
=o\{\widetilde\Delta_s(q)^2\},
\end{equation*}
for $u=2,3,4$,
\begin{equation*}
\sum_{\ell_1,\ldots,\ell_u\in\calL}
\left|\cum_q\{h^{(s)}_{\ell_1}(X_{[s]}),\ldots,h^{(s)}_{\ell_u}(X_{[s]})\}\right|
\le C_q\widetilde\Delta_s(q)^{u/2},
\end{equation*}
and, for $s>1$, for all $I_1,\ldots,I_4\in P_{qs}(n)$ such that every element of $I_1\cup\cdots\cup I_4$ appears in exactly two of the four index sets and $I_a\ne I_b$ for $a\ne b$,
\begin{equation}\label{eq:Lq_b2}
\E_0\left[
\sum_{\ell_1,\ldots,\ell_4\in\calL}
\prod_{a=1}^{4}(\otimes^q h^{(s)}_{\ell_a})(X_{I_a})
\right]
=o\{\widetilde\Delta_s(q)^2\}.
\end{equation}
\end{assumption}

\begin{remark}
Assumption \ref{ass:LqU} is a compact restatement of Assumptions 3.2--3.3 of \citet{ZhangWangShao2025} for the present component-wise independence indexing set $\calL=\Lam$.  It is used only to invoke the fixed-$L_q$ U-statistic CLT.  The new independence proof in this paper uses Assumptions \ref{ass:V}, \ref{ass:W} and \ref{ass:Q} to control the maximum statistic and the exceedance-event expansion.
\end{remark}

\subsection{Null calibration for finite $L_q$ statistics}\label{sec:null}

The following proposition gives the exact null centering and variance of \eqref{eq:lq_stat}.

\begin{proposition}\label{prop:mean_var_general}
Under $H_0$, for any fixed even $q\ge2$,
\begin{equation*}
\E_0(S_{A,q})=0,
\qquad
\Var_0(S_{A,q})=\npairs v_{A,q,n}.
\end{equation*}
Moreover, for fixed even $q_1,q_2$,
\begin{equation*}
\Cov_0(S_{A,q_1},S_{A,q_2})
=
\npairs\{\E_0(\wt A_{12}^{q_1+q_2})-
\E_0(\wt A_{12}^{q_1})\E_0(\wt A_{12}^{q_2})\}.
\end{equation*}
\end{proposition}

The next theorem gives the finite-$L_q$ null approximation.  The theorem is stated at the rank-correlation power-sum level.  For U-statistic classes, the proof uses the diagonal-free $L_q$ U-statistic theorem of \citet{ZhangWangShao2025} and the diagonal equivalence in Remark \ref{rem:powersum}.

\begin{theorem}\label{thm:lq_clt}
Let Assumption \ref{ass:A0} hold and let $\calQ=\{q_1,\ldots,q_K\}\subset2\mathbb Z^{+}$ be fixed.

\begin{enumerate}[label=(\roman*),leftmargin=0.35in]
\item If $A=V$ and Assumption \ref{ass:V} holds, then \eqref{eq:lq_clt_statement} below holds.
\item If $A=W$ and Assumptions \ref{ass:W} and \ref{ass:LqU} hold with $s=1$, then \eqref{eq:lq_clt_statement} below holds.
\item If $A=Q$ and Assumptions \ref{ass:Q} and \ref{ass:LqU} hold with $s=2$, then \eqref{eq:lq_clt_statement} below holds.
\end{enumerate}
Define
\[
\bm{Z}_{A,\calQ}=(Z_{A,q_1},\ldots,Z_{A,q_K})^\top.
\]
Then
\begin{equation}\label{eq:lq_clt_statement}
\bm{Z}_{A,\calQ}
\Rightarrow N_K(0,\mathbf{\Gamma}_{A,\calQ}),
\end{equation}
where
\begin{equation*}
(\mathbf{\Gamma}_{A,\calQ})_{jk}
=
\lim_{n\to\infty}
\frac{
\E_0(\wt A_{12}^{q_j+q_k})-
\E_0(\wt A_{12}^{q_j})\E_0(\wt A_{12}^{q_k})
}{
(v_{A,q_j,n}v_{A,q_k,n})^{1/2}
},
\end{equation*}
whenever the limits exist.  In particular, for every fixed even $q$,
\begin{equation*}
Z_{A,q}\Rightarrow N(0,1).
\end{equation*}
\end{theorem}

\begin{remark}
The finite-$L_q$ statistics are not generally mutually independent across different $q$ values when the power-sum version \eqref{eq:lq_stat} is used.  This is immaterial for the Cauchy combination test, which only requires valid marginal p-values and is robust to dependence among the finite-$L_q$ p-values.  The key independence result needed here is the block independence between the entire finite-$L_q$ vector and the maximum statistic.
\end{remark}

For finite $q$, define the one-sided p-value
\begin{equation}\label{eq:pval_finite}
P_{A,q}=1-\Phi(Z_{A,q}),
\qquad q\in2\mathbb Z^{+},
\end{equation}
where $\Phi$ is the standard normal distribution function.

\subsection{Maximum statistics and block asymptotic independence}\label{sec:independence}

For the simple linear and non-degenerate U-statistic classes, define
\begin{equation*}
L_A=\max_{(s,t)\in\Lam}|\wt A_{st}|,
\qquad A\in\{V,W\},
\end{equation*}
and let $\sigma_{A,n}^2=\Var_0(\wt A_{12})$.  The standardized maximum statistic is
\begin{equation*}
M_A=\frac{L_A^2}{\sigma_{A,n}^2}-4\log p+\log\log p,
\qquad A\in\{V,W\}.
\end{equation*}
Its limiting distribution function is
\begin{equation*}
G(y)=\exp\{-(8\pi)^{-1/2}\exp(-y/2)\}.
\end{equation*}

For the degenerate class, define
\begin{equation*}
L_Q=\max_{(s,t)\in\Lam}\wt Q_{h,st},
\end{equation*}
and
\begin{equation*}
M_Q=(n-1)\lambda_1^{-1}{m\choose2}^{-1}L_Q
-4\log p-(\mu_1-2)\log\log p+\Lambda/\lambda_1.
\end{equation*}
Its limiting distribution function is
\begin{equation*}
F(y)=\exp\left\{-\frac{2^{\mu_1/2-2}\kappa}{\Gamma(\mu_1/2)}\exp(-y/2)\right\},
\end{equation*}
where $\kappa$ is defined in \eqref{eq:kappa_general}.

\begin{theorem}\label{thm:block_independence}
Let $\calQ=\{q_1,\ldots,q_K\}\subset2\N$ be fixed.

\begin{enumerate}[label=(\roman*),leftmargin=0.35in]
\item If $A=V$, assume Assumptions \ref{ass:A0} and \ref{ass:V}.
\item If $A=W$, assume Assumptions \ref{ass:A0}, \ref{ass:W} and \ref{ass:LqU} with $s=1$.
\item If $A=Q$, assume Assumptions \ref{ass:A0}, \ref{ass:Q} and \ref{ass:LqU} with $s=2$.
\end{enumerate}
Then, for $A\in\{V,W\}$,
\begin{equation*}
(Z_{A,q_1},\ldots,Z_{A,q_K},M_A)
\Rightarrow
(Z_{A,q_1}^{(0)},\ldots,Z_{A,q_K}^{(0)},Y_A),
\end{equation*}
where $\bm{Z}_{A,\calQ}^{(0)}=(Z_{A,q_1}^{(0)},\ldots,Z_{A,q_K}^{(0)})^\top\sim N_K(0,\mathbf{\Gamma}_{A,\calQ})$, $Y_A$ has distribution function $G$, and the Gaussian vector is independent of $Y_A$.

For the degenerate class,
\begin{equation*}
(Z_{Q,q_1},\ldots,Z_{Q,q_K},M_Q)
\Rightarrow
(Z_{Q,q_1}^{(0)},\ldots,Z_{Q,q_K}^{(0)},Y_Q),
\end{equation*}
where $\bm{Z}_{Q,\calQ}^{(0)}=(Z_{Q,q_1}^{(0)},\ldots,Z_{Q,q_K}^{(0)})^\top\sim N_K(0,\mathbf{\Gamma}_{Q,\calQ})$, $Y_Q$ has distribution function $F$, and the Gaussian vector is independent of $Y_Q$.
\end{theorem}

Appendix \ref{app:block} proves the following rate form.  Let
\[
r_{A,\calQ}(n,p)=
\sup_{\bm{z}\in\mathbb R^K}
\left|\Pp_0\{(Z_{A,q_1},\ldots,Z_{A,q_K})\le \bm{z}\}-\Phi_{\mathbf{\Gamma}_{A,\calQ}}(\bm{z})\right|,
\]
where inequalities are componentwise.  For fixed $k\ge1$, $\varepsilon\in(0,1)$ and even integers $\tau_r\ge4$, $1\le r\le k$, the proof gives
\begin{align*}
&\left|\Pp_0\{D_p(\bm{x})\cap(M_A>y)\}-\Pp_0\{D_p(\bm{x})\}\Pp_0(M_A>y)\right| \\
&\quad \le
\sum_{r=1}^{k}
\left[\Delta_{p,\varepsilon}(\bm{x})H_A(p,r;y)
+C_{A,\calQ,r,\tau_r}\varepsilon^{-\tau_r}p^{2r-\tau_r/2}\right]
+2H_A(p,k+1;y),                                      
\end{align*}
where
\[
D_p(\bm{x})=\bigcap_{j=1}^{K}\{Z_{A,q_j}\le x_j\},
\quad
\Delta_{p,\varepsilon}(\bm{x})
\le 2r_{A,\calQ}(n,p)
+\sup_{\|\bm{u}-\bm{x}\|_\infty\le2\varepsilon}
|\Phi_{\mathbf{\Gamma}_{A,\calQ}}(\bm{u})-\Phi_{\mathbf{\Gamma}_{A,\calQ}}(\bm{x})|,
\]
and
\begin{equation*}
H_A(p,r;y)
\le
\frac{C_{A,y}^{r}}{2^r r!}
+2C_{A,y}^{r}p^{1-r}
+C_{A,y,r}(2r)^{2r}p^{-1}.
\end{equation*}
Taking $\tau_r=6r$, then $n,p\to\infty$, then $\varepsilon\downarrow0$, and finally $k\to\infty$ yields Theorem \ref{thm:block_independence}.

The maximum p-values are
\begin{equation}\label{eq:pmax}
P_{A,\infty}=1-G(M_A),\qquad A\in\{V,W\},
\end{equation}
for the simple linear and non-degenerate classes, and
\begin{equation}\label{eq:pmaxQ}
P_{Q,\infty}=1-F(M_Q)
\end{equation}
for the degenerate class.

\subsection{Cauchy combination of finite-$L_q$ and maximum p-values}\label{sec:cct}

Let
\begin{equation*}
\calB=\{2,4,6,\infty\}.
\end{equation*}
For a generic rank class $A\in\{V,W,Q\}$, collect p-values
\[
\{P_{A,a}:a\in\calB\},
\]
where finite-$q$ p-values are defined by \eqref{eq:pval_finite} and the maximum p-values are defined by \eqref{eq:pmax}--\eqref{eq:pmaxQ}.  Let $w_a\ge0$ and $\sum_{a\in\calB}w_a=1$.  Define
\begin{equation*}
\mathcal C_{A,\calB}=
\sum_{a\in\calB}w_a\tan\left[\pi\left\{\frac12-P_{A,a}\right\}\right],
\end{equation*}
and
\begin{equation*}
P_{A,\calB}
=\frac12-\frac1\pi\arctan(\mathcal C_{A,\calB}).
\end{equation*}
A default choice is
\begin{equation*}
w_2=w_4=w_6=w_\infty=\frac14.
\end{equation*}

\begin{theorem}\label{thm:cct}
Assume the conditions of Theorem \ref{thm:block_independence}.  Then the test
\[
\phi_{A,\calB,\alpha}=\ind\{P_{A,\calB}\le \alpha\}
\]
is asymptotically valid in the Cauchy tail sense:
\begin{equation}\label{eq:cct_level}
\lim_{\alpha\downarrow0}\limsup_{n,p\to\infty}
\left|
\frac{\Pp_0(P_{A,\calB}\le\alpha)}{\alpha}-1
\right|=0.
\end{equation}
Moreover, if any one of the constituent tests indexed by $a\in\calB$ is consistent against a sequence of alternatives, then
\[
\Pp_1(P_{A,\calB}\le\alpha)\to1.
\]
\end{theorem}

The Cauchy $L_{\calB}$ combination is useful here because $L_4$ and $L_6$ p-values may be strongly dependent with the $L_2$ p-value, while Theorem \ref{thm:block_independence} separates the finite-$L_q$ block from the maximum statistic.  The Cauchy transformation accommodates the remaining finite-block dependence without estimating or integrating the finite-$L_q$ Gaussian distribution.

\subsection{Representative examples: exact $L_4$ and $L_6$ centering and variance}\label{sec:examples}

This section gives exact finite-sample single-pair centering and variance formulas for the representative rank correlations.  These coefficients are examples only; the theory above is formulated for the three general rank-based classes.  Throughout this section,
\[
S_{T,q}=\sum_{(s,t)\in\Lam}\{T_{st}^q-\mu_{T,q,n}\},\qquad
Z_{T,q}=\frac{S_{T,q}}{\{\npairs v_{T,q,n}\}^{1/2}},
\]
where
\[
T_{st}=\wt A_{st},
\qquad
\mu_{T,q,n}=M_{T,q}(n),\qquad
v_{T,q,n}=M_{T,2q}(n)-M_{T,q}^2(n),
\qquad
M_{T,r}(n)=\E_0(T_{12}^r).
\]
Thus the global centering and variance are $\npairs\mu_{T,q,n}$ and $\npairs v_{T,q,n}$, respectively.  No limiting approximation is used in the formulas in this section.

\subsubsection{Spearman's $\rho$}\label{subsec:rho_exact}

Spearman's $\rho$ \citep{Spearman1904} belongs to the simple linear rank-statistic class.  Under $H_0$, fix the first rank sequence to be $(1,\ldots,n)$ and let the second rank sequence be a uniform permutation $\pi\in\Sf_n$.  With
\[
a_i=i-\frac{n+1}{2},\qquad
c_{\rho,n}=\frac{12}{n(n^2-1)},\qquad
\rho_n(\pi)=c_{\rho,n}\sum_{i=1}^{n}a_i a_{\pi(i)},
\]
we have $M_{\rho,r}(n)=n!^{-1}\sum_{\pi\in\Sf_n}\rho_n(\pi)^r$.  The partition identity
\begin{equation}\label{eq:rho_partition_moment_new}
M_{\rho,r}(n)
=c_{\rho,n}^r
\sum_{\lambda\vdash r}
\frac{r!}{\prod_{b=1}^{\ell(\lambda)}\lambda_b!\prod_{j\ge1}m_j(\lambda)!}
\frac{D_\lambda(n)^2}{\fall{n}{\ell(\lambda)}}
\end{equation}
gives exact symbolic moments.  In \eqref{eq:rho_partition_moment_new}, $\lambda\vdash r$ means that $\lambda=(\lambda_1,\ldots,\lambda_{\ell})$ is an integer partition of $r$, so $\lambda_1+\cdots+\lambda_{\ell}=r$ and $\lambda_1\ge\cdots\ge\lambda_{\ell}\ge1$; $\ell(\lambda)=\ell$ is the number of parts; $m_j(\lambda)=|\{b:\lambda_b=j\}|$ is the multiplicity of part $j$; the product $\prod_{j\ge1}m_j(\lambda)!$ is finite because all but finitely many $m_j(\lambda)$ are zero; and
\[
\fall{n}{\ell(\lambda)}=n(n-1)\cdots\{n-\ell(\lambda)+1\}
\]
is the falling factorial.  The symbol $\sum^{*}$ denotes summation over pairwise distinct indices.  Hence
\[
D_\lambda(n)=\sum_{i_1,\ldots,i_{\ell(\lambda)}=1}^{n\,*}\prod_{b=1}^{\ell(\lambda)}a_{i_b}^{\lambda_b}
\]
is the ordered distinct-index power sum associated with the partition $\lambda$.
Evaluating \eqref{eq:rho_partition_moment_new} gives the following closed forms:
\begin{align}
M_{\rho,4}(n)
&=
\frac{3(25n^3-38n^2-35n+72)}
{25n(n-1)^3(n+1)},\label{eq:rho_M4}\\
M_{\rho,6}(n)
&=
\frac{3P_{\rho,6}(n)}
{245n^3(n-1)^5(n+1)^3},\label{eq:rho_M6}
\end{align}
where
\begin{align*}
P_{\rho,6}(n)
={}&1225n^8-4361n^7-178n^6+23818n^5-22783n^4\\
&{}-50081n^3+54280n^2+44160n-28800 .
\end{align*}
The exact variances needed for the $L_4$ and $L_6$ statistics are
\begin{align}
v_{\rho,4,n}
&=
\frac{24(n-2)Q_{\rho,4}(n)}
{4375n^5(n-1)^7(n+1)^5},\label{eq:rho_v4}\\
v_{\rho,6,n}
&=
\frac{18(n-2)Q_{\rho,6}(n)}
{2789661875n^9(n-1)^{11}(n+1)^9},\label{eq:rho_v6}
\end{align}
with
\begin{align*}
Q_{\rho,4}(n)
={}&17500n^{12}-99575n^{11}+93952n^{10}+857943n^9\\
&{}-2236650n^8-3105081n^7+12836468n^6+8558537n^5\\
&{}-32726710n^4-20519664n^3+28279440n^2\\
&{}+9858240n-12700800,
\end{align*}
and
\begin{align*}
Q_{\rho,6}(n)
={}&1576158959375n^{22}-26956502698125n^{21}
+204016193881500n^{20}\\
&{}-656132617822682n^{19}-1171932384888603n^{18}
+16913917053629829n^{17}\\
&{}-33663135573263722n^{16}-143066811467638476n^{15}
+610987613264235129n^{14}\\
&{}+596842447834386253n^{13}-5189139972464602944n^{12}
-1409441833203864570n^{11}\\
&{}+27750786105920376371n^{10}+4444828679768649627n^9
-95698023681505100946n^8\\
&{}-19996738740525207104n^7+206938856876542180608n^6
+34634552355461373696n^5\\
&{}-313861911687028044288n^4-56314087053512122368n^3
+270499002102369976320n^2\\
&{}+37528151745373470720n-101439305560276992000 .
\end{align*}
Consequently,
\[
\mu_{\rho,4,n}=M_{\rho,4}(n),\qquad
\mu_{\rho,6,n}=M_{\rho,6}(n),
\]
and $v_{\rho,4,n}$ and $v_{\rho,6,n}$ are given in \eqref{eq:rho_v4} and \eqref{eq:rho_v6}.

\subsubsection{Kendall's $\tau$}\label{subsec:tau_exact}

Kendall's $\tau$ \citep{Kendall1938} belongs to the non-degenerate U-statistic class.  Let
\[
I(\pi)=\sum_{1\le i<j\le n}\ind\{\pi(i)>\pi(j)\}
\]
be the inversion number of a uniform permutation.  Kendall's statistic is
\[
\tau_n(\pi)=1-\frac{4I(\pi)}{n(n-1)}.
\]
The inversion number has the same law as $\sum_{j=1}^nU_j$, where $U_j$ are independent and uniform on $\{0,1,\ldots,j-1\}$.  Hence its cumulants are
\begin{equation*}
\kappa_r(I)=\frac{B_r^+}{r}\sum_{j=1}^{n}(j^r-1),
\end{equation*}
where $B_1^+=1/2$ and $B_r^+=B_r$ for $r\ne1$, and $B_r$ are the Bernoulli numbers.  Since $\E_0\tau_n=0$,
\[
\kappa_r(\tau_n)=
\left\{-\frac{4}{n(n-1)}\right\}^r\kappa_r(I),\qquad r\ge2.
\]
Using the complete Bell polynomial moment-cumulant formula gives the exact moments.  In simplified form,
\begin{align}
M_{\tau,4}(n)
&=
\frac{4P_{\tau,4}(n)}
{675n^3(n-1)^3},\label{eq:tau_M4}\\
M_{\tau,6}(n)
&=
\frac{8P_{\tau,6}(n)}
{59535n^5(n-1)^5},\label{eq:tau_M6}
\end{align}
where
\[
P_{\tau,4}(n)
=
100n^4+328n^3-127n^2-997n-372,
\]
and
\begin{align*}
P_{\tau,6}(n)
={}&9800n^7+32732n^6-42010n^5-230695n^4\\
&{}-72460n^3+400733n^2+391500n+118080 .
\end{align*}
The exact $L_4$ and $L_6$ variance constants are
\begin{align}
v_{\tau,4,n}
&=
\frac{256(n-2)Q_{\tau,4}(n)}
{9568125n^7(n-1)^7},\label{eq:tau_v4}\\
v_{\tau,6,n}
&=
\frac{128(n-2)Q_{\tau,6}(n)}
{164726744056875n^{11}(n-1)^{11}},\label{eq:tau_v6}
\end{align}
where
\begin{align*}
Q_{\tau,4}(n)
={}&140000n^9+617400n^8-160764n^7-4827762n^6\\
&{}-7764663n^5+3028185n^4+23170684n^3\\
&{}+31403277n^2+20222343n+5273100,
\end{align*}
and
\begin{align*}
Q_{\tau,6}(n)
={}&100874173400000n^{15}+106587400920000n^{14}
-1350735059674000n^{13}\\
&{}-2750236703502288n^{12}+705156071105876n^{11}
+18114848707300164n^{10}\\
&{}+74210935173807565n^9+32519698879088181n^8
-379954037364238322n^7\\
&{}-639273543932846298n^6+136412983449767425n^5
+1193679769717739457n^4\\
&{}+1569782012721160896n^3+1559385642899802384n^2\\
&{}+1047269150681247360n+285343922116915200 .
\end{align*}
Thus $\mu_{\tau,4,n}=M_{\tau,4}(n)$, $\mu_{\tau,6,n}=M_{\tau,6}(n)$, and the variance constants are \eqref{eq:tau_v4} and \eqref{eq:tau_v6}.

\subsubsection{Degenerate examples and finite-sample calibration}\label{subsec:deg_exact_examples}

Hoeffding's $D$ \citep{Hoeffding1948b}, Blum--Kiefer--Rosenblatt's $R$ \citep{BlumKieferRosenblatt1961} and Bergsma--Dassios--Yanagimoto's $\tau^\ast$ \citep{BergsmaDassios2014} are representative members of the degenerate rank-based U-statistic class.  The exact finite-sample moments required by the $L_4$ and $L_6$ versions involve fourth-, sixth-, eighth- and twelfth-order products of degenerate rank kernels.  Their evaluation requires a high-order overlap enumeration over rank patterns.  Appendix \ref{app:omega} gives the coefficient-extraction formula and prints the full exact $L_4$ calculation for $\tau^\ast$; this already illustrates the size of the rational arrays involved.

For the finite-sample implementation of $D$, $R$ and $\tau^\ast$, we therefore use high-precision null simulation at each fixed sample size $n$.  Let $B=1,000,000$ and let $T_n^{(1)},\ldots,T_n^{(B)}$ be iid null values of the single-pair statistic generated from independent random permutations under $H_0$.  For $q\in\{4,6\}$ define
\begin{equation}\label{eq:deg_mc_muvar}
\widehat\mu_{T,q,n}^{\rm MC}=B^{-1}\sum_{b=1}^{B}\{T_n^{(b)}\}^{q},
\qquad
\widehat v_{T,q,n}^{\rm MC}=B^{-1}\sum_{b=1}^{B}\{T_n^{(b)}\}^{2q}
-\{\widehat\mu_{T,q,n}^{\rm MC}\}^2 .
\end{equation}
The implemented finite-$L_q$ statistic for $T\in\{D,R,\tau^\ast\}$ is
\begin{equation}\label{eq:deg_mc_Z}
Z_{T,q}^{\rm MC}
=
\frac{\sum_{1\le s<t\le p}T_{st}^{q}-\npairs\widehat\mu_{T,q,n}^{\rm MC}}
{\{\npairs\widehat v_{T,q,n}^{\rm MC}\}^{1/2}},
\qquad q\in\{4,6\}.
\end{equation}
This calibration is used for the size and power simulations reported below.  For substantially larger $n$, the leading analytic constants in Appendix \ref{app:leading} provide a fast approximation and a useful check on the simulation-based calibration.

\begin{table}[!htbp]
\centering
\caption{Finite-sample centering and variance used by the representative finite-$L_q$ statistics. Spearman's $\rho$ and Kendall's $\tau$ use closed rational formulas.  The degenerate rank correlations use the fixed-$n$ Monte Carlo calibration in \eqref{eq:deg_mc_muvar}; Appendix \ref{app:omega} gives exact overlap formulas and the complete $\tau^\ast$ $L_4$ example.}
\label{tab:exact_examples_new}
\begin{tabular}{lcccc}
\toprule
$T$ & $\mu_{T,4,n}$ & $v_{T,4,n}$ & $\mu_{T,6,n}$ & $v_{T,6,n}$ \\
\midrule
$\rho$ & \eqref{eq:rho_M4} & \eqref{eq:rho_v4} & \eqref{eq:rho_M6} & \eqref{eq:rho_v6}\\
$\tau$ & \eqref{eq:tau_M4} & \eqref{eq:tau_v4} & \eqref{eq:tau_M6} & \eqref{eq:tau_v6}\\
$D$ & \multicolumn{4}{c}{Monte Carlo calibration \eqref{eq:deg_mc_muvar} with $B=1,000,000$}\\
$R$ & \multicolumn{4}{c}{Monte Carlo calibration \eqref{eq:deg_mc_muvar} with $B=1,000,000$}\\
$\tau^\ast$ & \multicolumn{4}{c}{Monte Carlo calibration \eqref{eq:deg_mc_muvar}; exact $L_4$ formula in Appendix \ref{app:omega}}\\
\bottomrule
\end{tabular}
\end{table}

\section{Simulation studies}\label{sec:simulation}

This section describes the simulation design used to evaluate the proposed finite-$L_q$ rank-based tests.  The goal is to compare the empirical size and power of the component tests based on $L_2,L_4,L_6$ and $L_\infty$, and of the $L_{2,4,6,\infty}$ combination, across dense, moderately sparse and very sparse alternatives.

For each data-generating mechanism, we generate $N_{\mathrm{rep}}=1000$ Monte Carlo samples and report rejection frequencies at nominal level $\alpha=0.05$.  The dimensions are chosen from
\[
(n,p)\in\{(100,100),(100,200),(200,200),(200,400)\}.
\]
For each representative rank coefficient
\[
T\in\{\rho,\tau,D,R,\tau^\ast\},
\]
we compute the four p-values
\[
P_{T,2},\qquad P_{T,4},\qquad P_{T,6},\qquad P_{T,\infty},
\]
and the equal-weight combined p-value
\[
P_{T,\{2,4,6,\infty\}}
=\frac12-\frac1\pi\arctan\left
\{
\frac14\sum_{a\in\{2,4,6,\infty\}}
\tan\bigl[\pi\{1/2-P_{T,a}\}\bigr]
\right\}.
\]
For Spearman's $\rho$ and Kendall's $\tau$, the finite-$L_4$ and finite-$L_6$ rank statistics are centered and scaled by the exact formulas in Section \ref{sec:examples}.  For $D$, $R$ and $\tau^\ast$, the same statistics use the fixed-$n$ null Monte Carlo calibration \eqref{eq:deg_mc_muvar} with $B=1,000,000$.  As a non-rank benchmark, we also report the same family of tests constructed from Pearson's sample correlation coefficient.  The Pearson benchmark is included only for comparison and is not covered by the rank-based theory.

Under the null hypothesis, the coordinates are mutually independent and the marginal distributions are generated from three models:
\[
\mathrm{N}(0,1),
\qquad
 t_3/\sqrt{3},
\qquad
(\chi^2_1-1)/\sqrt{2}.
\]
The last two distributions are included to assess the robustness of the rank-based procedures under heavy tails and skewness.

\subsection*{Empirical size under the null}

Tables \ref{tab:size_pearson}--\ref{tab:size_tau_star} report the empirical size results under the three null marginal distributions.  All entries are rejection frequencies multiplied by 100.  The nominal level is 5\%, and each entry is based on 1000 Monte Carlo replications.  The column $L_{2,4,6,\infty}$ combines $L_2,L_4,L_6$ and $L_\infty$ by the equal-weight Cauchy rule, whereas $L_{2,\infty}$ combines only $L_2$ and $L_\infty$ and corresponds to the original max-sum comparison.  For $D$, $R$ and $\tau^\ast$, the finite-$L_4$ and finite-$L_6$ calibrations use fixed-$n$ null Monte Carlo moments with $B=1,000,000$, because using only the leading asymptotic constants tends to inflate the null rejection probabilities in moderate samples.

\begin{table}[!htbp]
\centering
\scriptsize
\caption{Empirical size for Pearson's sample correlation coefficient. Entries are empirical rejection frequencies multiplied by 100.}
\label{tab:size_pearson}
\begin{tabular}{ccrrrrrr}
\toprule
$(n,p)$ & Marginal law & $L_2$ & $L_4$ & $L_6$ & $L_\infty$ & $L_{2,4,6,\infty}$ & $L_{2,\infty}$ \\
\midrule
$(100,100)$ & Normal & 5.5 & 5.4 & 5.5 & 1.1 & 4.5 & 3.3 \\
$(100,100)$ & $t_3/\sqrt{3}$ & 7.5 & 87.8 & 97.6 & 76.2 & 95.7 & 67.9 \\
$(100,100)$ & $(\chi^2_1-1)/\sqrt{2}$ & 9.3 & 94.5 & 99.8 & 81.0 & 99.6 & 68.2 \\
\addlinespace
$(100,200)$ & Normal & 5.2 & 5.5 & 6.3 & 1.8 & 5.4 & 2.6 \\
$(100,200)$ & $t_3/\sqrt{3}$ & 8.9 & 100.0 & 100.0 & 97.6 & 100.0 & 94.0 \\
$(100,200)$ & $(\chi^2_1-1)/\sqrt{2}$ & 9.4 & 100.0 & 100.0 & 95.5 & 100.0 & 88.3 \\
\addlinespace
$(200,200)$ & Normal & 3.3 & 3.5 & 5.1 & 1.8 & 3.4 & 2.5 \\
$(200,200)$ & $t_3/\sqrt{3}$ & 8.3 & 99.9 & 100.0 & 98.4 & 100.0 & 97.5 \\
$(200,200)$ & $(\chi^2_1-1)/\sqrt{2}$ & 5.9 & 100.0 & 100.0 & 91.5 & 100.0 & 81.9 \\
\addlinespace
$(200,400)$ & Normal & 5.2 & 5.5 & 7.0 & 2.2 & 4.6 & 3.3 \\
$(200,400)$ & $t_3/\sqrt{3}$ & 7.3 & 100.0 & 100.0 & 100.0 & 100.0 & 100.0 \\
$(200,400)$ & $(\chi^2_1-1)/\sqrt{2}$ & 6.8 & 100.0 & 100.0 & 99.4 & 100.0 & 97.4 \\
\bottomrule
\end{tabular}
\end{table}

\begin{table}[!htbp]
\centering
\scriptsize
\caption{Empirical size for Spearman's $\rho$. Entries are empirical rejection frequencies multiplied by 100.}
\label{tab:size_rho}
\begin{tabular}{ccrrrrrr}
\toprule
$(n,p)$ & Marginal law & $L_2$ & $L_4$ & $L_6$ & $L_\infty$ & $L_{2,4,6,\infty}$ & $L_{2,\infty}$ \\
\midrule
$(100,100)$ & Normal & 4.6 & 5.0 & 5.6 & 1.2 & 4.6 & 2.9 \\
$(100,100)$ & $t_3/\sqrt{3}$ & 4.8 & 4.6 & 4.5 & 1.4 & 4.6 & 3.7 \\
$(100,100)$ & $(\chi^2_1-1)/\sqrt{2}$ & 6.0 & 6.7 & 7.0 & 3.8 & 7.0 & 5.9 \\
\addlinespace
$(100,200)$ & Normal & 5.1 & 5.7 & 6.1 & 2.2 & 5.5 & 3.4 \\
$(100,200)$ & $t_3/\sqrt{3}$ & 5.6 & 6.3 & 6.3 & 1.1 & 5.1 & 3.5 \\
$(100,200)$ & $(\chi^2_1-1)/\sqrt{2}$ & 4.3 & 4.6 & 4.1 & 1.7 & 4.6 & 2.7 \\
\addlinespace
$(200,200)$ & Normal & 4.3 & 4.3 & 5.3 & 2.6 & 4.2 & 3.4 \\
$(200,200)$ & $t_3/\sqrt{3}$ & 5.1 & 4.9 & 5.3 & 3.2 & 4.8 & 4.6 \\
$(200,200)$ & $(\chi^2_1-1)/\sqrt{2}$ & 5.4 & 5.2 & 5.9 & 3.5 & 5.3 & 4.4 \\
\addlinespace
$(200,400)$ & Normal & 5.2 & 6.2 & 5.9 & 2.8 & 5.2 & 3.6 \\
$(200,400)$ & $t_3/\sqrt{3}$ & 4.7 & 6.2 & 6.6 & 1.7 & 5.2 & 3.2 \\
$(200,400)$ & $(\chi^2_1-1)/\sqrt{2}$ & 5.5 & 5.2 & 5.7 & 2.3 & 5.1 & 3.2 \\
\bottomrule
\end{tabular}
\end{table}

\begin{table}[!htbp]
\centering
\scriptsize
\caption{Empirical size for Kendall's $\tau$. Entries are empirical rejection frequencies multiplied by 100.}
\label{tab:size_tau}
\begin{tabular}{ccrrrrrr}
\toprule
$(n,p)$ & Marginal law & $L_2$ & $L_4$ & $L_6$ & $L_\infty$ & $L_{2,4,6,\infty}$ & $L_{2,\infty}$ \\
\midrule
$(100,100)$ & Normal & 4.6 & 5.0 & 5.2 & 2.3 & 5.2 & 3.5 \\
$(100,100)$ & $t_3/\sqrt{3}$ & 5.4 & 4.3 & 4.5 & 2.3 & 4.6 & 4.2 \\
$(100,100)$ & $(\chi^2_1-1)/\sqrt{2}$ & 6.0 & 6.2 & 7.9 & 4.0 & 7.3 & 6.5 \\
\addlinespace
$(100,200)$ & Normal & 4.9 & 5.5 & 6.1 & 3.8 & 5.5 & 4.3 \\
$(100,200)$ & $t_3/\sqrt{3}$ & 5.3 & 6.1 & 5.5 & 2.0 & 4.6 & 3.8 \\
$(100,200)$ & $(\chi^2_1-1)/\sqrt{2}$ & 4.2 & 5.0 & 4.9 & 3.2 & 5.0 & 3.2 \\
\addlinespace
$(200,200)$ & Normal & 4.3 & 4.0 & 5.2 & 3.9 & 4.4 & 3.8 \\
$(200,200)$ & $t_3/\sqrt{3}$ & 5.0 & 5.2 & 5.2 & 5.2 & 5.7 & 5.2 \\
$(200,200)$ & $(\chi^2_1-1)/\sqrt{2}$ & 5.2 & 4.9 & 6.2 & 4.6 & 6.0 & 5.3 \\
\addlinespace
$(200,400)$ & Normal & 5.3 & 6.0 & 5.8 & 4.0 & 5.8 & 4.2 \\
$(200,400)$ & $t_3/\sqrt{3}$ & 4.5 & 6.1 & 6.6 & 2.9 & 5.5 & 3.7 \\
$(200,400)$ & $(\chi^2_1-1)/\sqrt{2}$ & 5.5 & 5.0 & 5.5 & 4.0 & 5.0 & 3.7 \\
\bottomrule
\end{tabular}
\end{table}

\begin{table}[!htbp]
\centering
\scriptsize
\caption{Empirical size for Hoeffding's $D$. Entries are empirical rejection frequencies multiplied by 100.}
\label{tab:size_D}
\begin{tabular}{ccrrrrrr}
\toprule
$(n,p)$ & Marginal law & $L_2$ & $L_4$ & $L_6$ & $L_\infty$ & $L_{2,4,6,\infty}$ & $L_{2,\infty}$ \\
\midrule
$(100,100)$ & Normal & 7.1 & 6.1 & 1.7 & 6.1 & 5.6 & 6.9 \\
$(100,100)$ & $t_3/\sqrt{3}$ & 5.3 & 4.3 & 2.0 & 5.6 & 4.4 & 5.8 \\
$(100,100)$ & $(\chi^2_1-1)/\sqrt{2}$ & 8.0 & 7.8 & 3.6 & 7.4 & 8.1 & 9.8 \\
\addlinespace
$(100,200)$ & Normal & 7.6 & 8.1 & 4.4 & 8.8 & 8.6 & 8.9 \\
$(100,200)$ & $t_3/\sqrt{3}$ & 7.5 & 7.3 & 2.5 & 7.2 & 6.4 & 7.7 \\
$(100,200)$ & $(\chi^2_1-1)/\sqrt{2}$ & 7.1 & 7.2 & 3.4 & 8.8 & 7.2 & 8.8 \\
\addlinespace
$(200,200)$ & Normal & 4.2 & 5.6 & 5.3 & 6.4 & 6.2 & 5.6 \\
$(200,200)$ & $t_3/\sqrt{3}$ & 4.3 & 5.9 & 6.0 & 6.6 & 7.7 & 7.2 \\
$(200,200)$ & $(\chi^2_1-1)/\sqrt{2}$ & 5.5 & 6.7 & 5.8 & 7.2 & 7.3 & 7.0 \\
\addlinespace
$(200,400)$ & Normal & 5.2 & 5.8 & 5.9 & 7.0 & 7.4 & 7.7 \\
$(200,400)$ & $t_3/\sqrt{3}$ & 4.8 & 6.5 & 7.5 & 8.0 & 7.3 & 5.7 \\
$(200,400)$ & $(\chi^2_1-1)/\sqrt{2}$ & 4.2 & 5.6 & 6.0 & 5.8 & 6.2 & 4.6 \\
\bottomrule
\end{tabular}
\end{table}

\begin{table}[!htbp]
\centering
\scriptsize
\caption{Empirical size for Blum--Kiefer--Rosenblatt's $R$. Entries are empirical rejection frequencies multiplied by 100.}
\label{tab:size_R}
\begin{tabular}{ccrrrrrr}
\toprule
$(n,p)$ & Marginal law & $L_2$ & $L_4$ & $L_6$ & $L_\infty$ & $L_{2,4,6,\infty}$ & $L_{2,\infty}$ \\
\midrule
$(100,100)$ & Normal & 6.4 & 6.0 & 2.1 & 2.8 & 5.6 & 5.3 \\
$(100,100)$ & $t_3/\sqrt{3}$ & 4.9 & 4.7 & 2.5 & 2.8 & 4.5 & 4.6 \\
$(100,100)$ & $(\chi^2_1-1)/\sqrt{2}$ & 7.2 & 8.7 & 3.9 & 4.9 & 8.1 & 7.6 \\
\addlinespace
$(100,200)$ & Normal & 7.2 & 9.4 & 4.9 & 3.9 & 7.9 & 6.9 \\
$(100,200)$ & $t_3/\sqrt{3}$ & 7.4 & 8.1 & 4.1 & 3.6 & 7.2 & 6.1 \\
$(100,200)$ & $(\chi^2_1-1)/\sqrt{2}$ & 6.1 & 7.3 & 3.5 & 3.4 & 6.0 & 4.9 \\
\addlinespace
$(200,200)$ & Normal & 4.8 & 4.9 & 3.7 & 4.5 & 4.8 & 4.3 \\
$(200,200)$ & $t_3/\sqrt{3}$ & 3.9 & 5.0 & 4.0 & 4.2 & 5.2 & 4.9 \\
$(200,200)$ & $(\chi^2_1-1)/\sqrt{2}$ & 4.7 & 5.6 & 4.7 & 5.1 & 5.4 & 5.4 \\
\addlinespace
$(200,400)$ & Normal & 5.2 & 4.6 & 4.3 & 3.9 & 5.4 & 5.5 \\
$(200,400)$ & $t_3/\sqrt{3}$ & 4.8 & 5.6 & 4.4 & 3.7 & 5.3 & 3.2 \\
$(200,400)$ & $(\chi^2_1-1)/\sqrt{2}$ & 4.2 & 4.4 & 3.8 & 3.5 & 4.4 & 2.9 \\
\bottomrule
\end{tabular}
\end{table}

\begin{table}[!htbp]
\centering
\scriptsize
\caption{Empirical size for Bergsma--Dassios--Yanagimoto's $\tau^\ast$. Entries are empirical rejection frequencies multiplied by 100.}
\label{tab:size_tau_star}
\begin{tabular}{ccrrrrrr}
\toprule
$(n,p)$ & Marginal law & $L_2$ & $L_4$ & $L_6$ & $L_\infty$ & $L_{2,4,6,\infty}$ & $L_{2,\infty}$ \\
\midrule
$(100,100)$ & Normal & 6.8 & 6.0 & 1.7 & 3.0 & 5.7 & 5.4 \\
$(100,100)$ & $t_3/\sqrt{3}$ & 5.4 & 4.2 & 2.3 & 3.6 & 4.4 & 4.9 \\
$(100,100)$ & $(\chi^2_1-1)/\sqrt{2}$ & 7.9 & 8.3 & 3.8 & 5.6 & 7.8 & 7.8 \\
\addlinespace
$(100,200)$ & Normal & 7.5 & 9.2 & 4.6 & 5.3 & 7.7 & 6.8 \\
$(100,200)$ & $t_3/\sqrt{3}$ & 7.3 & 7.5 & 3.0 & 4.0 & 6.7 & 5.9 \\
$(100,200)$ & $(\chi^2_1-1)/\sqrt{2}$ & 6.6 & 7.0 & 3.5 & 4.1 & 6.2 & 5.3 \\
\addlinespace
$(200,200)$ & Normal & 4.2 & 5.1 & 4.1 & 4.8 & 5.3 & 4.3 \\
$(200,200)$ & $t_3/\sqrt{3}$ & 3.8 & 4.9 & 4.7 & 5.0 & 6.4 & 5.7 \\
$(200,200)$ & $(\chi^2_1-1)/\sqrt{2}$ & 4.8 & 6.3 & 5.2 & 5.6 & 6.1 & 5.7 \\
\addlinespace
$(200,400)$ & Normal & 4.9 & 5.0 & 4.8 & 4.7 & 5.9 & 5.7 \\
$(200,400)$ & $t_3/\sqrt{3}$ & 5.0 & 5.9 & 5.2 & 4.3 & 6.3 & 3.4 \\
$(200,400)$ & $(\chi^2_1-1)/\sqrt{2}$ & 4.2 & 4.5 & 4.1 & 4.0 & 5.3 & 3.3 \\
\bottomrule
\end{tabular}
\end{table}

The empirical size patterns can be read in three groups.  First, the Pearson benchmark is close to the nominal level under Gaussian margins, but it becomes severely oversized under heavy-tailed and skewed margins, especially for $L_4$, $L_6$, $L_\infty$ and their combined tests.  This confirms that high-order Pearson-power statistics are highly sensitive to marginal nonnormality.  Second, for the simple linear and non-degenerate rank coefficients, namely Spearman's $\rho$ and Kendall's $\tau$, the finite-$L_q$ procedures are generally close to the 5\% nominal level, while the $L_\infty$ component is conservative in several moderate-dimensional settings.  Third, for the degenerate rank coefficients $D$, $R$ and $\tau^\ast$, the $n=100$ cases show mild upward size distortion for some $L_2$, $L_4$ and $L_{2,4,6,\infty}$ entries, but the calibration improves substantially when $n=200$.  The degenerate rank tables use fixed-$n$ null Monte Carlo moments with $B=1,000,000$; Appendix \ref{app:omega} gives the complete exact $\tau^\ast$ $L_4$ calculation as a check on the combinatorial complexity.

\subsection*{Empirical power under sparse-to-dense alternatives}

We next examine the finite-sample power of the proposed tests under sparse-to-dense alternatives. In this experiment,
\[
  n=100,\qquad p=200,\qquad \alpha=0.05,
\]
Each configuration is based on \(1000\) Monte Carlo replications. The sparsity parameter is \(k\in\{2,\ldots,16\}\). For each replication, an active set of \(k\) coordinates is selected uniformly at random. Conditional on the active set, the latent Gaussian vector \(\bm{Z}=(Z_1,\ldots,Z_p)^\top\) has unit marginal variances and pairwise correlations only inside the active block. The nonzero correlations are generated independently from \(U(r_{\min},r_{\max})\), where
\[
\begin{aligned}
&\text{(i) linear normal:}&&
  r_{\min}=\{12k^{-1}\log(p)/n\}^{1/2},\quad
  r_{\max}=\{14k^{-1}\log(p)/n\}^{1/2},\\
&\text{(ii) sine cube-root:}&&
  r_{\min}=\{9\log(p)/(n\log((k+2)/2))\}^{1/2},\\
&&& r_{\max}=\{10\log(p)/(n\log((k+2)/2))\}^{1/2},\\
&\text{(iii) sine cubic:}&&
  r_{\min}=\{19k^{-1/2}\log(p)/n\}^{1/2},\quad
  r_{\max}=\{20k^{-1/2}\log(p)/n\}^{1/2}.
\end{aligned}
\]
The observed vector \(\bm{X}=(X_1,\ldots,X_p)^\top\) is then generated from one of the following three models:
\[
  X_j=Z_j,\qquad
  X_j=\sin\{2\pi\,\operatorname{sgn}(Z_j)|Z_j|^{1/3}/3\},\qquad
  X_j=\sin(\pi Z_j^3/4).
\]

We compare six pairwise dependence measures: Pearson's \(r\), Kendall's \(\tau\), Spearman's \(\rho\), the Bergsma--Dassios--Yanagimoto \(\tau^\ast\), Hoeffding's \(D\), and the Blum--Kiefer--Rosenblatt \(R\). For each dependence measure, we report tests based on \(L_2\), \(L_4\), \(L_6\), \(L_\infty\), the \(L_{2,\infty}\) combination and the \(L_{2,4,6,\infty}\) combination. Because the higher-order Pearson tests have inflated size under non-Gaussian null distributions, Pearson's \(r\) is reported only for the \(L_2\) statistic in the power comparison. In the coefficient-level comparison, Pearson's \(r\) uses \(L_2\), whereas the five rank-based dependence measures use the \(L_{2,4,6,\infty}\) combination.

\begin{figure}[!htbp]
  \centering
  \includegraphics[width=\textwidth]{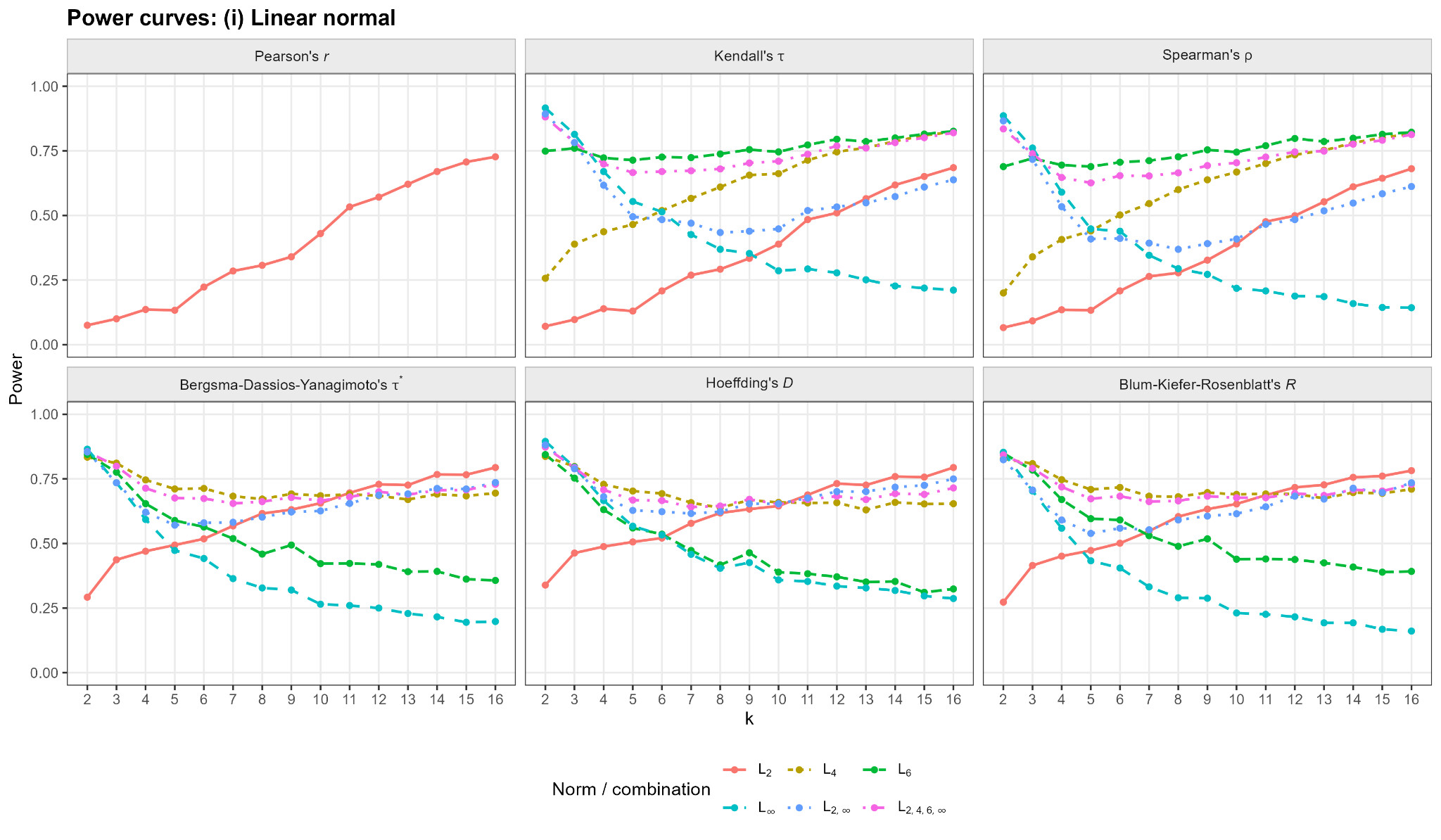}
  \caption{Power curves for the linear normal alternative. Each panel corresponds to one dependence measure. Pearson's \(r\) is reported only for \(L_2\).}
  \label{fig:power-linear}
\end{figure}

\begin{figure}[!htbp]
  \centering
  \includegraphics[width=\textwidth]{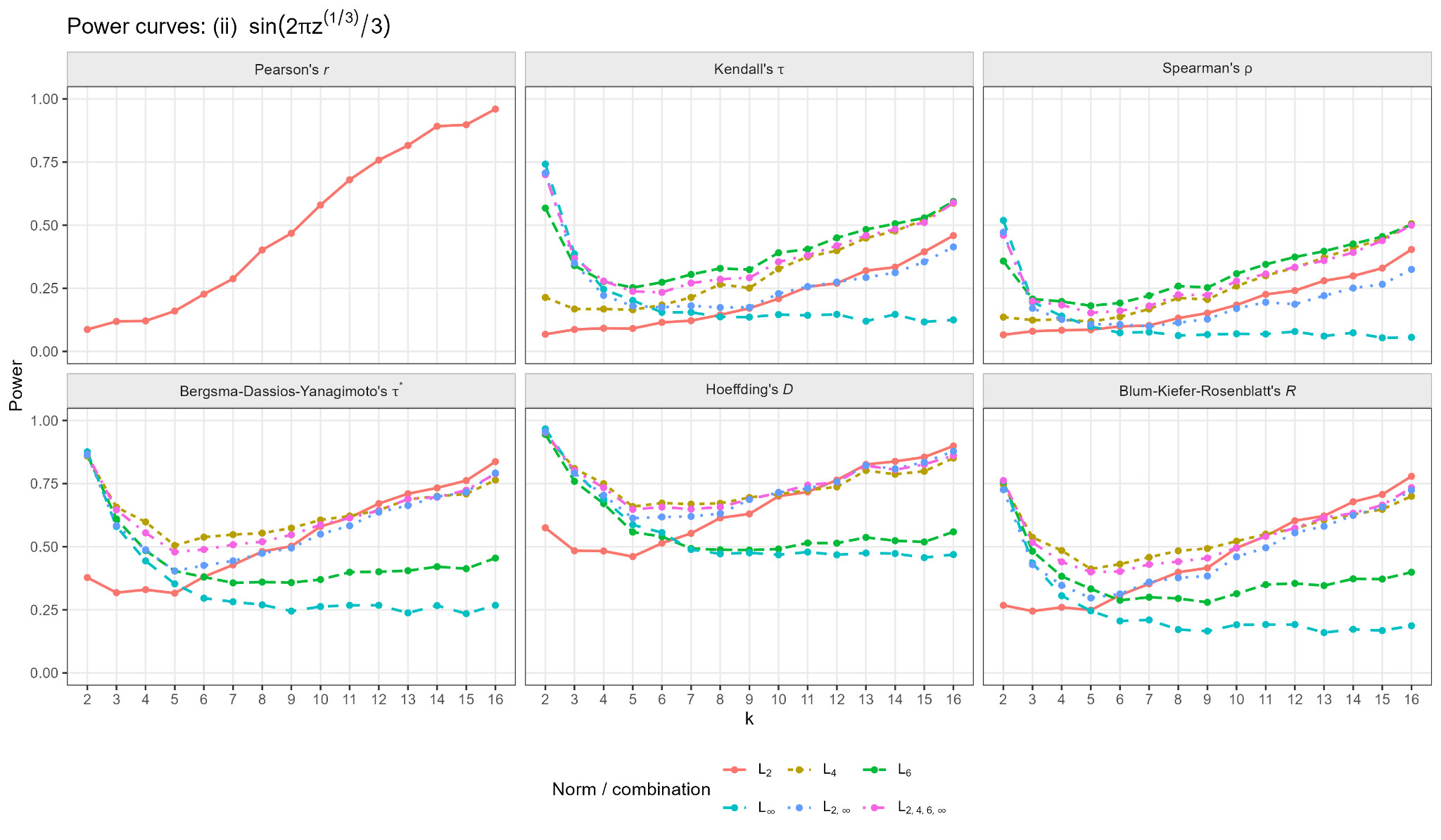}
  \caption{Power curves for the sine cube-root alternative. Each panel corresponds to one dependence measure. Pearson's \(r\) is reported only for \(L_2\).}
  \label{fig:power-sin-cuberoot}
\end{figure}

\begin{figure}[!htbp]
  \centering
  \includegraphics[width=\textwidth]{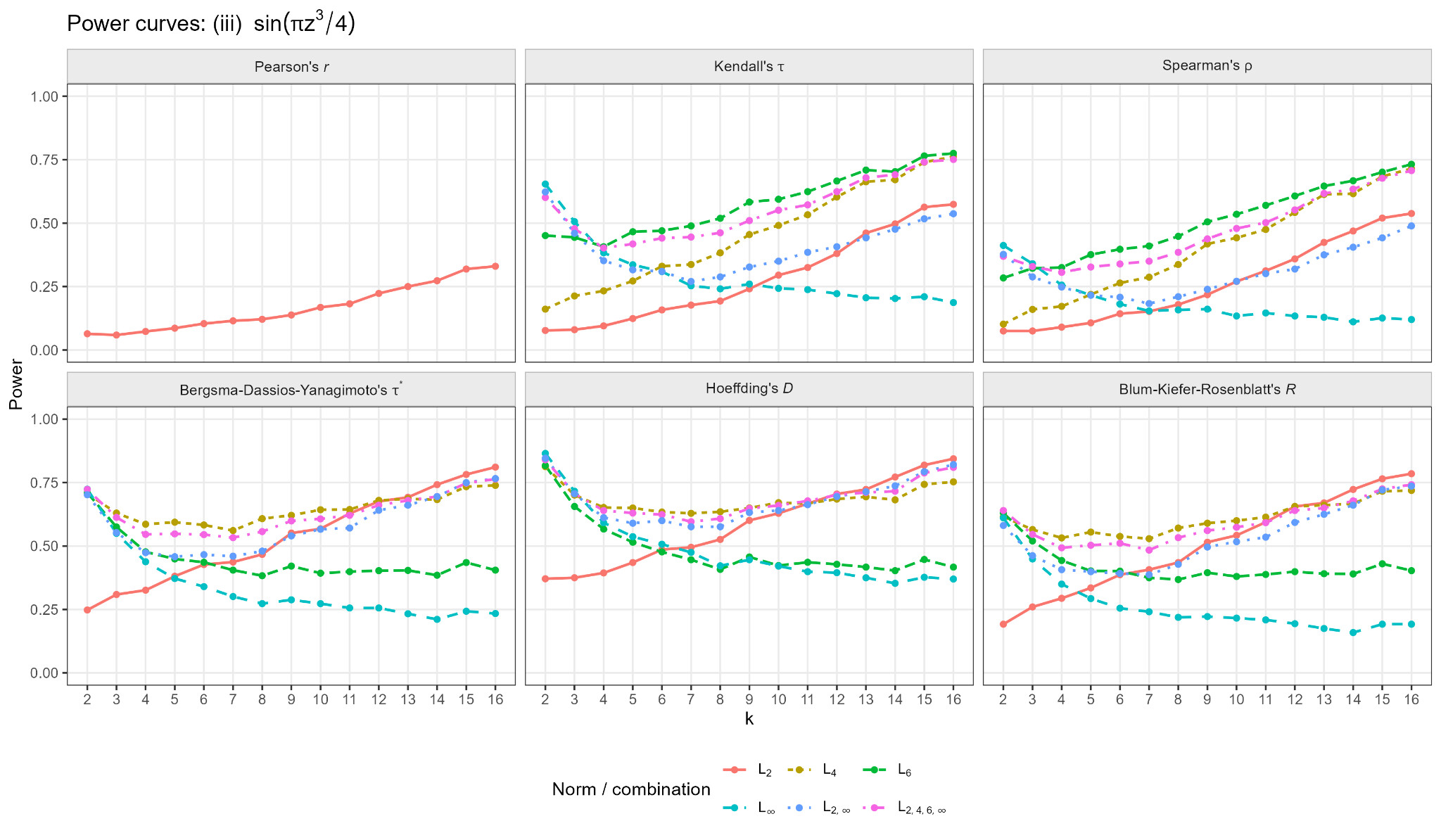}
  \caption{Power curves for the sine cubic alternative. Each panel corresponds to one dependence measure. Pearson's \(r\) is reported only for \(L_2\).}
  \label{fig:power-sin-cube}
\end{figure}

\begin{figure}[!htbp]
  \centering
  \includegraphics[width=\textwidth]{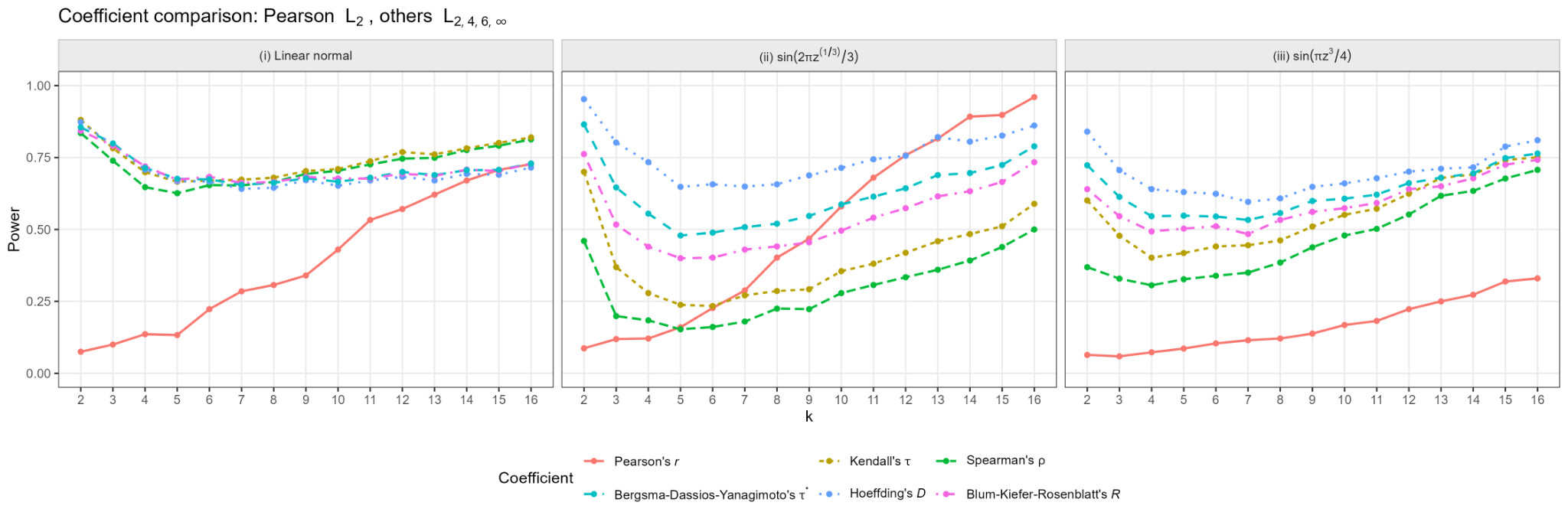}
  \caption{Comparison of dependence measures across the three alternatives. Pearson's \(r\) uses \(L_2\), whereas Kendall's \(\tau\), Spearman's \(\rho\), \(\tau^\ast\), Hoeffding's \(D\), and Blum--Kiefer--Rosenblatt's \(R\) use \(L_{2,4,6,\infty}\).}
  \label{fig:power-coef-comparison}
\end{figure}

Figure~\ref{fig:power-linear} shows the results for the linear normal alternative. The \(L_\infty\) and high-order \(L_6\) tests are most powerful when the alternative is very sparse. For example, for Hoeffding's \(D\), \(L_\infty\) has power \(0.895\) at \(k=2\), but decreases to \(0.287\) at \(k=16\); its \(L_6\) test decreases from \(0.844\) to \(0.324\). In contrast, the \(L_2\) statistic becomes stronger as the signal becomes less sparse, increasing from \(0.339\) to \(0.794\) for Hoeffding's \(D\). The combined tests provide a compromise over the sparsity range. For Pearson's \(r\), the \(L_2\) power increases from \(0.075\) at \(k=2\) to \(0.727\) at \(k=16\), reflecting the fact that \(L_2\) is better matched to dense alternatives.

Figure~\ref{fig:power-sin-cuberoot} presents a more nonlinear alternative. Pearson's \(L_2\) is highly effective in this setting, increasing from \(0.087\) at \(k=2\) to \(0.960\) at \(k=16\). Among the rank-based and nonparametric measures, Hoeffding's \(D\) and \(\tau^\ast\) are the most stable. Their \(L_{2,4,6,\infty}\) powers remain high at \(k=16\), with values \(0.861\) and \(0.789\), respectively. Kendall's \(\tau\) and Spearman's \(\rho\) are less powerful for this nonlinear transformation, especially at intermediate sparsity levels, but their power improves again as \(k\) increases.

Figure~\ref{fig:power-sin-cube} gives the results for the sine cubic alternative. Pearson's \(L_2\) is weak in this setting, increasing only from \(0.064\) to \(0.330\). The nonparametric measures are much more competitive. At \(k=16\), the \(L_{2,4,6,\infty}\) powers are \(0.810\) for Hoeffding's \(D\), \(0.764\) for \(\tau^\ast\), \(0.742\) for Blum--Kiefer--Rosenblatt's \(R\), \(0.751\) for Kendall's \(\tau\), and \(0.707\) for Spearman's \(\rho\). As in the linear setting, the high-order tests for \(D\), \(R\), and \(\tau^\ast\) are strongest under sparse signals and become less favorable when \(k\) grows; for instance, the \(L_6\) power for \(D\) decreases from \(0.817\) at \(k=2\) to \(0.417\) at \(k=16\). The \(L_2\) tests move in the opposite direction and are better for denser alternatives.

The coefficient-level comparison in Figure~\ref{fig:power-coef-comparison} summarizes the main conclusions. First, no single dependence measure dominates uniformly. Pearson's \(r\) can be very strong for monotone or nearly linear dense alternatives, but it is much less effective for the sine cubic alternative. Second, Hoeffding's \(D\) and \(\tau^\ast\) provide robust power across all three alternatives and are particularly effective for nonlinear dependence. Third, the \(L_{2,4,6,\infty}\) combination adapts well across different sparsity regimes for the nonparametric dependence measures, although the individual \(L_q\) components reveal the expected tradeoff: \(L_\infty\) and high-order finite norms favor sparse alternatives, while \(L_2\) favors dense alternatives. Finally, the raw \(L_4\) and \(L_6\) statistics for Kendall's \(\tau\) and Spearman's \(\rho\) may not decrease with \(k\) as sharply as those for \(D\), \(R\), and \(\tau^\ast\), because even powers of centered correlation-type statistics still contain lower-order signal components under alternatives.
\FloatBarrier

\section{Discussion}\label{sec:discussion}

We propose a finite-$L_q$ extension of rank-based mutual independence testing and prove that a fixed finite-$L_q$ block is asymptotically independent of the corresponding maximum statistic.  The resulting $L_{2,4,6,\infty}$ procedure combines distribution-free rank robustness with stable power over dense, moderately sparse and sparse alternatives.

Future work includes developing faster exact finite-sample calibration for high-order degenerate rank U-statistics, possibly using symbolic enumeration and dynamic programming ideas related to those in \citet{DrtonHanShi2020} and \citet{ZhangWangShao2025}.  Another direction is to extend finite-$L_q$ rank combinations to conditional, residual-based and kernel or distance-based independence settings; relevant starting points include \citet{WangLiuFengMa2024}, \citet{YaoZhangShao2018} and \citet{PfisterEtAl2018}.

\appendix

\section{Finite-sample moment calculation for degenerate rank correlations}\label{app:omega}

This appendix describes the exact finite-sample moment calculation for the degenerate rank correlations Hoeffding's $D$, Blum--Kiefer--Rosenblatt's $R$ and Bergsma--Dassios--Yanagimoto's $\tau^\ast$.  The formulas are written at the level of concrete rank kernels.  The full $L_4$ centering and variance calculation is printed for $\tau^\ast$ because it is the shortest of the three degenerate examples and already shows the combinatorial complexity of exact finite-$n$ calibration.

For $m\in\{4,5,6\}$, let $\vct e_m=(1,\ldots,m)$ and let $\Sf_m$ be the permutation group on $m$ elements.  For $u=(u_1,\ldots,u_m)$ and $\omega\in\Sf_m$, define
\begin{align*}
A_{m,a}(u;\omega)
&=
\{\ind(u_{\omega_1}\le u_{\omega_a})-\ind(u_{\omega_2}\le u_{\omega_a})\}
\{\ind(u_{\omega_3}\le u_{\omega_a})-\ind(u_{\omega_4}\le u_{\omega_a})\},\\
B_4(u;\omega)
&=
\ind(u_{\omega_1},u_{\omega_3}<u_{\omega_2},u_{\omega_4})
+\ind(u_{\omega_2},u_{\omega_4}<u_{\omega_1},u_{\omega_3})\\
&\quad-
\ind(u_{\omega_1},u_{\omega_4}<u_{\omega_2},u_{\omega_3})
-
\ind(u_{\omega_2},u_{\omega_3}<u_{\omega_1},u_{\omega_4}),
\end{align*}
where $\ind(a,b<c,d)=\ind(a<c)\ind(a<d)\ind(b<c)\ind(b<d)$.  The three pattern kernels are
\begin{align}
\psi_D(\sigma)
&=
\frac1{16}
\sum_{\omega\in\Sf_5}
A_{5,5}(\vct e_5;\omega)A_{5,5}(\sigma;\omega),
\qquad \sigma\in\Sf_5, \label{eq:psi_D_explicit}\\
\psi_R(\sigma)
&=
\frac1{32}
\sum_{\omega\in\Sf_6}
A_{6,5}(\vct e_6;\omega)A_{6,6}(\sigma;\omega),
\qquad \sigma\in\Sf_6, \label{eq:psi_R_explicit}\\
\psi_{\tau^\ast}(\sigma)
&=
\frac1{4!}
\sum_{\omega\in\Sf_4}
B_4(\vct e_4;\omega)B_4(\sigma;\omega),
\qquad \sigma\in\Sf_4. \label{eq:psi_taustar_explicit}
\end{align}
Equivalently, the pattern-value tables are
\[
\begin{array}{c|c|c}
\toprule
T & \psi_T(\sigma) & \#\{\sigma\in\Sf_{m_T}:\psi_T(\sigma)\}\\
\midrule
D & 1 & 8\\
D & 0 & 96\\
D & -1/2 & 16\\
\midrule
R & 1 & 80\\
R & 1/2 & 128\\
R & 0 & 64\\
R & -1/4 & 320\\
R & -1/2 & 128\\
\midrule
\tau^\ast & 2/3 & 8\\
\tau^\ast & -1/3 & 16\\
\bottomrule
\end{array}
\]
Under $H_0$, the joint ranks of one coordinate pair are equivalent to $(1,\pi_1),\ldots,(n,\pi_n)$ with $\pi$ uniform on $\Sf_n$.  Therefore
\begin{equation}\label{eq:T_pattern_stat}
T_n(\pi)
=
{n\choose m_T}^{-1}
\sum_{I\in { [n]\choose m_T}}
\psi_T\{\operatorname{pat}_I(\pi)\},
\end{equation}
where ${[n]\choose m_T}=\{A\subseteq[n]: |A|=m_T\}$; for $I=\{i_1<\cdots<i_m\}$, $\operatorname{pat}_I(\pi)\in\Sf_m$ is the relative rank pattern of $(\pi_{i_1},\ldots,\pi_{i_m})$.

For integers $m,r,b$, let
\[
\mathcal I_{m,r}(b)
=
\left\{
(I_1,\ldots,I_r): I_a\in {[b]\choose m},\;1\le a\le r,\;
\bigcup_{a=1}^{r}I_a=[b]
\right\}.
\]
For $T\in\{D,R,\tau^\ast\}$, define
\begin{equation}\label{eq:Omega_def}
\Omega_{T,r,b}
=
\frac1{b!}
\sum_{\pi\in\Sf_b}
\sum_{(I_1,\ldots,I_r)\in\mathcal I_{m_T,r}(b)}
\prod_{a=1}^{r}
\psi_T\{\operatorname{pat}_{I_a}(\pi)\},
\qquad m_T\le b\le r m_T .
\end{equation}
For $T\in\{D,R,\tau^\ast\}$ define
\begin{equation}\label{eq:Pr_def_main}
\mathsf P_{T,r}(n)=\sum_{b=m_T}^{r m_T}{n\choose b}\Omega_{T,r,b},
\qquad {n\choose b}=0\quad(b>n).
\end{equation}
Then the exact single-pair moment is
\begin{equation}\label{eq:deg_M_binomial}
M_{T,r}(n)=\E_0(T_n^r)={n\choose m_T}^{-r}\mathsf P_{T,r}(n).
\end{equation}
Indeed, expanding $T_n^r$ and grouping the ordered kernel tuples according to the cardinality of their union gives \eqref{eq:Omega_def} and hence \eqref{eq:deg_M_binomial}.  Equivalently, each summand can be written as an explicit rational function of $n$:
\begin{equation}\label{eq:A_T_r_b_term}
\frac{{n\choose b}}{{n\choose m_T}^{r}}\Omega_{T,r,b}
=
\Omega_{T,r,b}\frac{(m_T!)^r}{b!}
\frac{\prod_{j=0}^{b-1}(n-j)}{\{\prod_{j=0}^{m_T-1}(n-j)\}^{r}}.
\end{equation}

Substituting \eqref{eq:psi_D_explicit}--\eqref{eq:psi_taustar_explicit} gives the coefficient-level formulas
\begin{align}
\Omega_{D,r,b}
&=
\frac{16^{-r}}{b!}
\sum_{\pi\in\Sf_b}
\sum_{(I_1,\ldots,I_r)\in\mathcal I_{5,r}(b)}
\prod_{a=1}^{r}
\sum_{\omega_a\in\Sf_5}
A_{5,5}(\vct e_5;\omega_a)
A_{5,5}\{\operatorname{pat}_{I_a}(\pi);\omega_a\},
\label{eq:Omega_D_formula}\\
\Omega_{R,r,b}
&=
\frac{32^{-r}}{b!}
\sum_{\pi\in\Sf_b}
\sum_{(I_1,\ldots,I_r)\in\mathcal I_{6,r}(b)}
\prod_{a=1}^{r}
\sum_{\omega_a\in\Sf_6}
A_{6,5}(\vct e_6;\omega_a)
A_{6,6}\{\operatorname{pat}_{I_a}(\pi);\omega_a\},
\label{eq:Omega_R_formula}\\
\Omega_{\tau^\ast,r,b}
&=
\frac{(4!)^{-r}}{b!}
\sum_{\pi\in\Sf_b}
\sum_{(I_1,\ldots,I_r)\in\mathcal I_{4,r}(b)}
\prod_{a=1}^{r}
\sum_{\omega_a\in\Sf_4}
B_4(\vct e_4;\omega_a)
B_4\{\operatorname{pat}_{I_a}(\pi);\omega_a\}.
\label{eq:Omega_tau_formula}
\end{align}
Equations \eqref{eq:Omega_D_formula}--\eqref{eq:Omega_tau_formula} are the exact rational-constant calculations used to form $\mathsf P_{T,r}(n)$.  The second-order constants, which verify the normalization against known formulas, are
\begin{align*}
&\Omega_{D,2,5}=\frac1{10},\qquad
\Omega_{D,2,6}=\frac{41}{45},\qquad
\Omega_{D,2,7}=\frac{49}{30},\qquad
\Omega_{D,2,8}=\frac{28}{45},\\
&\Omega_{D,2,9}=\Omega_{D,2,10}=0,\\[1mm]
&\Omega_{R,2,6}=\frac{41}{180},
\quad
\Omega_{R,2,7}=\frac{287}{60},
\quad
\Omega_{R,2,8}=\frac{952}{45},
\quad
\Omega_{R,2,9}=\frac{154}{5},
\quad
\Omega_{R,2,10}=14,\\
&\Omega_{R,2,11}=\Omega_{R,2,12}=0,\\[1mm]
&\Omega_{\tau^\ast,2,4}=\frac29,
\quad
\Omega_{\tau^\ast,2,5}=\frac{32}{45},
\quad
\Omega_{\tau^\ast,2,6}=\frac25,
\quad
\Omega_{\tau^\ast,2,7}=\Omega_{\tau^\ast,2,8}=0.
\end{align*}

For $\tau^\ast$, let $C_n$ be the number of concordant quadruples in a uniform permutation of $\{1,\ldots,n\}$ and set
\[
K_n=3C_n-{n\choose4}=3{n\choose4}\tau_n^\ast .
\]
For $s=4$ and $s=8$,
\[
\E_0(K_n^s)=\sum_{b=4}^{3s}{n\choose b}a_{s,b},
\qquad
\Omega_{\tau^\ast,s,b}=3^{-s}a_{s,b}.
\]
Consequently, the exact $L_4$ centering and variance of the single-pair $\tau^\ast$ statistic are
\begin{equation}\label{eq:taustar_app_l4_exact}
\mu_{\tau^\ast,4,n}
=
\frac{\sum_{b=4}^{12}{n\choose b}a_{4,b}}{3^4{n\choose4}^4},
\qquad
v_{\tau^\ast,4,n}
=
\frac{\sum_{b=4}^{24}{n\choose b}a_{8,b}}{3^8{n\choose4}^8}
-\left\{\frac{\sum_{b=4}^{12}{n\choose b}a_{4,b}}{3^4{n\choose4}^4}\right\}^2 .
\end{equation}
The case $s=4$ gives the centering of $(\tau_n^\ast)^4$; the case $s=8$ gives the second moment of $(\tau_n^\ast)^4$ and hence its variance.  The complete exact rational arrays are as follows.

\begingroup
\small
\begin{longtable}{r>{\raggedright\arraybackslash}p{0.78\textwidth}}
\caption{Exact coefficient array for $s=4$ in the BDY $\tau^\ast$ $L_4$ calibration.  The listed coefficients satisfy $a_{4,b}=3^4\Omega_{\tau^\ast,4,b}$.}\label{tab:taustar_a4_full}\\
\toprule
$b$ & $a_{4,b}$\\
\midrule
\endfirsthead
\toprule
$b$ & $a_{4,b}$\\
\midrule
\endhead
4 & \texttt{6}\\
5 & \texttt{3984/5}\\
6 & \texttt{86562/5}\\
7 & \texttt{955356/7}\\
8 & \texttt{35817807/70}\\
9 & \texttt{71271603/70}\\
10 & \texttt{193676346/175}\\
11 & \texttt{108817236/175}\\
12 & \texttt{24762672/175}\\
\bottomrule
\end{longtable}
\endgroup

\begingroup
\small
\begin{longtable}{r>{\raggedright\arraybackslash}p{0.78\textwidth}}
\caption{Exact coefficient array for $s=8$ in the BDY $\tau^\ast$ $L_4$ calibration.  The listed coefficients satisfy $a_{8,b}=3^8\Omega_{\tau^\ast,8,b}$.}\label{tab:taustar_a8_full}\\
\toprule
$b$ & $a_{8,b}$\\
\midrule
\endfirsthead
\toprule
$b$ & $a_{8,b}$\\
\midrule
\endhead
4 & \texttt{86}\\
5 & \texttt{33679664/5}\\
6 & \texttt{45496828962/5}\\
7 & \texttt{2122691100468}\\
8 & \texttt{1646180207747391/10}\\
9 & \texttt{58827843848226249/10}\\
10 & \texttt{2909587818455523588/25}\\
11 & \texttt{392391897688632740043/275}\\
12 & \texttt{3209024638270591317216/275}\\
13 & \texttt{47841695172441710032608/715}\\
14 & \texttt{1393300388395409948078292/5005}\\
15 & \texttt{1655608772922828585830064/1925}\\
16 & \texttt{28632270526546652173175859/14300}\\
17 & \texttt{50601484615962860792463843/14300}\\
18 & \texttt{2020522303337817247422479631/425425}\\
19 & \texttt{4089733503008671502965210851/850850}\\
20 & \texttt{30687508392318266320595051829/8508500}\\
21 & \texttt{2364007315483141042586664069/1215500}\\
22 & \texttt{19681985307484680384868128/27625}\\
23 & \texttt{51591815442807446448342/325}\\
24 & \texttt{448521135302439977278464/27625}\\
\bottomrule
\end{longtable}
\endgroup

For reproducibility, the exact arithmetic used for the coefficient enumeration is equivalently summarized by the following coefficient-polynomial identity:
\begin{equation}\label{eq:omega_repro_identity}
\mathsf P_{T,r}(n)
=
\E_0\left[
\left\{
\sum_{I\in{[n]\choose m_T}}
\psi_T\{\operatorname{pat}_I(\Pi)\}
\right\}^{r}
\right],
\qquad \Pi\sim \operatorname{Unif}(\Sf_n),
\end{equation}
expanded in the binomial basis $\{{n\choose b}:m_T\le b\le r m_T\}$.  The coefficient of ${n\choose b}$ in this expansion is exactly $\Omega_{T,r,b}$.  Thus the calculation of the high-order constants reduces to a finite rational expansion in the binomial basis and is independent of $p$.

\section{Proof of Theorem \ref{thm:lq_clt}}\label{app:lqclt}

For $I=(s,t)\in\Lam$ put
\[
Y_{I,q}=\wt A_I^{\,q}-\mu_{A,q,n},
\qquad
S_{A,q}=\sum_{I\in\Lam}Y_{I,q}.
\]
Then
\begin{equation}\label{eq:app_E_zero}
\E_0S_{A,q}=\sum_{I\in\Lam}\E_0Y_{I,q}=0.
\end{equation}
For $I=(s,t)$ and $J=(u,v)$,
\begin{equation}\label{eq:app_cov_decomp}
\Cov_0(S_{A,q_1},S_{A,q_2})
=
\sum_{I,J\in\Lam}\E_0(Y_{I,q_1}Y_{J,q_2}).
\end{equation}
If $I\cap J=\varnothing$, then
\begin{equation*}
\E_0(Y_{I,q_1}Y_{J,q_2})=
\E_0Y_{I,q_1}\E_0Y_{J,q_2}=0.
\end{equation*}
If $I\ne J$ and $I\cap J\ne\varnothing$, write, without loss of generality, $I=(s,t)$ and $J=(s,u)$. Let
\[
\mathcal R_s=\sigma(R_{n1}^{s},\ldots,R_{nn}^{s}).
\]
Under $H_0$, the rank vectors in coordinates $t$ and $u$ are conditionally independent given $\mathcal R_s$; moreover, rank invariance gives
\begin{equation*}
\E_0(\wt A_{st}^{q}\mid\mathcal R_s)=\E_0(\wt A_{st}^{q})=\mu_{A,q,n},
\qquad q\in2\mathbb N.
\end{equation*}
Therefore
\begin{align}
\E_0(Y_{st,q_1}Y_{su,q_2})
&=
\E_0\left[\E_0\{Y_{st,q_1}Y_{su,q_2}\mid\mathcal R_s\}\right]       \notag\\
&=
\E_0\left[\E_0\{Y_{st,q_1}\mid\mathcal R_s\}
          \E_0\{Y_{su,q_2}\mid\mathcal R_s\}\right]                 \notag\\
&=0.                                                     \label{eq:app_overlap_cov}
\end{align}
Combining \eqref{eq:app_cov_decomp}--\eqref{eq:app_overlap_cov},
\begin{align}
\Cov_0(S_{A,q_1},S_{A,q_2})
&=
\sum_{I\in\Lam}\E_0(Y_{I,q_1}Y_{I,q_2})                                \notag\\
&=
\npairs\{\E_0(\wt A_{12}^{q_1+q_2})
       -\E_0(\wt A_{12}^{q_1})\E_0(\wt A_{12}^{q_2})\}.       \label{eq:app_cov_final}
\end{align}
Equations \eqref{eq:app_E_zero} and \eqref{eq:app_cov_final} prove Proposition \ref{prop:mean_var_general}.

Let
\[
r_{A,\calQ}(n,p)=
\sup_{\bm{z}\in\mathbb R^K}
\left|
\Pp_0\{(Z_{A,q_1},\ldots,Z_{A,q_K})\le \bm{z}\}
-\Phi_{\mathbf{\Gamma}_{A,\calQ}}(\bm{z})
\right|.
\]
For the U-statistic classes, the connection between the power statistic $S_{A,q}$ and the diagonal-free statistic used in the general $L_q$ U-statistic theory is as follows.  Write a symmetric U-statistic as
\[
A_\ell=\fall{n}{m}^{-1}\sum_{i_1,\ldots,i_m}^{*}
h_\ell(\bm X_{i_1},\ldots,\bm X_{i_m}),
\qquad \ell\in\calL=\Lam,
\]
where $\sum^*$ is over mutually distinct indices.  Let
\[
\mathcal J_{q,m}=\{(c,d):1\le c\le q,\;1\le d\le m\}.
\]
Denote by $\mathfrak P_{q,m}$ the finite set of partitions $\pi$ of $\mathcal J_{q,m}$ such that two positions $(c,d_1)$ and $(c,d_2)$ from the same copy $c$ are never in the same block.  Let $b(\pi)$ be the number of blocks of $\pi$.  If $\beta_\pi(c,d)$ is the block label containing $(c,d)$, define the collapsed kernel
\[
H_{\ell,\pi}(z_1,\ldots,z_{b(\pi)})
=
\prod_{c=1}^{q}
h_\ell\{z_{\beta_\pi(c,1)},\ldots,z_{\beta_\pi(c,m)}\}
\]
and its diagonal-free U-statistic
\[
U_{\ell,\pi}
=
\fall{n}{b(\pi)}^{-1}
\sum_{i_1,\ldots,i_{b(\pi)}}^{*}
H_{\ell,\pi}(\bm X_{i_1},\ldots,\bm X_{i_{b(\pi)}}).
\]
Grouping the terms in $A_\ell^q$ according to their equality pattern gives the exact identity
\[
A_\ell^q
=
\sum_{\pi\in\mathfrak P_{q,m}}
c_\pi(n)U_{\ell,\pi},
\qquad
c_\pi(n)=\frac{\fall{n}{b(\pi)}}{\fall{n}{m}^q}.
\]
The all-distinct partition $\pi_0$ has $b(\pi_0)=qm$ and gives the usual diagonal-free statistic
\[
U_{n,q}=\sum_{\ell\in\calL}U_{\ell,\pi_0}
=
\sum_{\ell\in\calL}\fall{n}{qm}^{-1}
\sum_{i_1,\ldots,i_{qm}}^{*}
\prod_{c=1}^{q}
 h_\ell(\bm X_{i_{(c-1)m+1}},\ldots,\bm X_{i_{cm}}).
\]
Therefore the coefficient multiplying $U_{n,q}$ in the exact power expansion is
\[
c_{A,q,n}=c_{\pi_0}(n)=\frac{\fall{n}{qm}}{\fall{n}{m}^{q}}
=1-\frac{q(q-1)m^2}{2n}+O(n^{-2}),
\]
so $c_{A,q,n}$ is deterministic, does not depend on $p$, and is bounded away from zero and infinity for fixed $q$ and $m$.  The identity for the centered statistic is
\[
S_{A,q}
=
c_{A,q,n}U_{n,q}
+
\sum_{\pi\in\mathfrak P_{q,m}\setminus\{\pi_0\}}
c_\pi(n)
\sum_{\ell\in\calL}
\{U_{\ell,\pi}-\E_0U_{\ell,\pi}\}.
\]
Thus $S_{A,q}$ is not approximated by a scalar multiple of the single all-distinct statistic alone when higher-order diagonal patterns have the same stochastic order.  For $q=4$ and $q=6$, some diagonal patterns may contribute to the limiting covariance, and they must be kept in the leading expansion.

Let $r_\pi$ be the degeneracy order of the centered collapsed kernel $H_{\ell,\pi}-\E_0H_{\ell,\pi}$, and set
\[
\mathfrak P_{q,m}^{\rm lead}
=
\{\pi\in\mathfrak P_{q,m}: r_\pi+2(qm-b(\pi))=qs\},
\]
where $s$ is the degeneracy order of $h_\ell$.  Define
\[
S_{A,q}^{\rm lead}
=
\sum_{\pi\in\mathfrak P_{q,m}^{\rm lead}}
c_\pi(n)
\sum_{\ell\in\calL}\{U_{\ell,\pi}-\E_0U_{\ell,\pi}\},
\qquad
R_{A,q}^{\rm rem}=S_{A,q}-S_{A,q}^{\rm lead}.
\]
Because $q$ and $m$ are fixed, $\mathfrak P_{q,m}$ is finite.  For a partition $\pi$, the factor $c_\pi(n)$ is of order $n^{b(\pi)-qm}$ and a bounded U-statistic with degeneracy order $r_\pi$ has variance of order $n^{-r_\pi}$.  Hence every non-leading partition has at least one additional power of $n^{-1}$ relative to the leading variance $n^{-qs}$.  More precisely, under Assumption \ref{ass:LqU},
\[
\Var_0(R_{A,q}^{\rm rem})
\le
C_{A,q}\npairs n^{-qs}
\delta_{A,q,n},
\]
where
\[
\delta_{A,q,n}
=
 n^{-1}
+
\sum_{t=s+1}^{m}
\frac{\Delta_t(q)}{n^{q(t-s)}\Delta_s(q)}.
\]
Moreover $\npairs v_{A,q,n}\asymp\npairs n^{-qs}$ for the fixed even orders considered here.  Consequently,
\[
\rho_{A,q,n}
=
\left\{
\frac{\Var_0(R_{A,q}^{\rm rem})}{\npairs v_{A,q,n}}
\right\}^{1/2}
\le
C_{A,q}
\left[
 n^{-1}
+
\sum_{t=s+1}^{m}
\frac{\Delta_t(q)}{n^{q(t-s)}\Delta_s(q)}
\right]^{1/2},
\]
and
\[
D_{A,\calQ}(n,p)
:=
\max_{q\in\calQ}
\left|
\frac{R_{A,q}^{\rm rem}}{(\npairs v_{A,q,n})^{1/2}}
\right|
=
O_{P_0}\left(
\max_{q\in\calQ}\rho_{A,q,n}
\right).
\]
By \eqref{eq:Lq_leading}, $\max_{q\in\calQ}\rho_{A,q,n}\to0$ and therefore $D_{A,\calQ}(n,p)\to0$ in probability.

The vector $(S_{A,q}^{\rm lead}:q\in\calQ)$ is a finite linear combination of diagonal-free U-statistics whose kernels are the collapsed kernels $H_{\ell,\pi}$ with $\pi\in\mathfrak P_{q,m}^{\rm lead}$.  Applying Theorem 3.2 of \citet{ZhangWangShao2025} to this finite collection of kernels and then using the Cramer--Wold device gives
\begin{equation}\label{eq:app_JASA_limit}
\sup_{\bm{z}\in\mathbb R^K}
\left|
\Pp_0\left\{
\left(
\frac{S_{A,q_1}^{\rm lead}}{(\npairs v_{A,q_1,n})^{1/2}},\ldots,
\frac{S_{A,q_K}^{\rm lead}}{(\npairs v_{A,q_K,n})^{1/2}}
\right)
\le \bm{z}
\right\}
-\Phi_{\mathbf{\Gamma}_{A,\calQ}}(\bm{z})
\right|
\le R_{U,\calQ}(n,p),
\qquad R_{U,\calQ}(n,p)\to0.
\end{equation}
Combining \eqref{eq:app_JASA_limit} with the bound for $D_{A,\calQ}(n,p)$ yields, for every $\varepsilon>0$,
\[
r_{A,\calQ}(n,p)
\le R_{U,\calQ}(n,p)
+
\Pp_0\{D_{A,\calQ}(n,p)>\varepsilon\}
+
\sup_{\|\bm{u}-\bm{z}\|_\infty\le\varepsilon}
|\Phi_{\mathbf{\Gamma}_{A,\calQ}}(\bm{u})-\Phi_{\mathbf{\Gamma}_{A,\calQ}}(\bm{z})|.
\]
Thus $r_{A,\calQ}(n,p)\to0$ for $A=W,Q$.

For the simple linear rank class, set
\[
Y_{st,q}=\wt V_{st}^{q}-\mu_{V,q,n},
\qquad
\sigma_{q,n}^2=\npairs v_{V,q,n}.
\]
For each fixed $r\ge2$ and fixed even $q$,
\begin{equation}\label{eq:app_linear_moment_bound}
\E_0|Y_{12,q}|^r
\le C_{q,r} n^{-qr/2},
\qquad
c_q n^{-q}\le v_{V,q,n}\le C_q n^{-q}.
\end{equation}
For $\bm{a}=(a_1,\ldots,a_K)^\top$ define
\[
\xi_{t}(\bm{a})=
\sum_{j=1}^{K}a_j\sigma_{q_j,n}^{-1}
\sum_{s=1}^{t-1}Y_{st,q_j},
\qquad
\mathcal F_t=\sigma(\bm{X}_{\cdot1},\ldots,\bm{X}_{\cdot t}).
\]
Then
\begin{equation*}
\E_0\{\xi_t(\bm{a})\mid\mathcal F_{t-1}\}=0,
\qquad
\sum_{t=2}^{p}\xi_t(\bm{a})=\sum_{j=1}^{K}a_jZ_{V,q_j}.
\end{equation*}
To compute the conditional variance, put
\[
\gamma_{jk,n}=\E_0(Y_{12,q_j}Y_{12,q_k}),
\qquad
\Gamma_{V,\calQ,n,jk}=\frac{\gamma_{jk,n}}{(v_{V,q_j,n}v_{V,q_k,n})^{1/2}}.
\]
For $s<r$, define
\[
\eta_{sr}^{jk}
=
\E_0\{Y_{s t,q_j}Y_{r t,q_k}
       \mid \mathcal R_s,\mathcal R_r\},
\qquad t>r,
\]
where the right-hand side does not depend on the particular value of $t$ by exchangeability.  Expanding the square of $\xi_t(\bm a)$ gives
\begin{align}
\E_0\{\xi_t(\bm a)^2\mid\mathcal F_{t-1}\}
&=
\sum_{j,k=1}^{K}
\frac{a_ja_k}{\sigma_{q_j,n}\sigma_{q_k,n}}
\left\{(t-1)\gamma_{jk,n}
+2\sum_{1\le s<r<t}\eta_{sr}^{jk}\right\}.\label{eq:app_condvar_expand}
\end{align}
Consequently,
\begin{align}
B_p(\bm a)
&:=\sum_{t=2}^{p}\E_0\{\xi_t(\bm a)^2\mid\mathcal F_{t-1}\}                                      \notag\\
&=
\sum_{j,k=1}^{K}
\frac{a_ja_k}{\sigma_{q_j,n}\sigma_{q_k,n}}
\left\{\npairs\gamma_{jk,n}
+2\sum_{1\le s<r\le p}(p-r)\eta_{sr}^{jk}\right\}.\label{eq:app_B_decomp}
\end{align}
The permutation moment calculation behind \eqref{eq:app_linear_moment_bound} also gives
\begin{equation}\label{eq:app_eta_moment}
\E_0\eta_{12}^{jk}=0,
\qquad
\E_0|\eta_{12}^{jk}|^2
\le C_{jk}n^{-q_j-q_k-1}.
\end{equation}
Indeed, after conditioning on two rank sequences, the conditional covariance is a finite sum of products of centered finite-population inner products.  The basic identity
\[
\E_0\left(\sum_{i=1}^{n}u_i v_{\pi(i)}\right)^2
=\frac{1}{n-1}\left(\sum_{i=1}^{n}u_i^2\right)
                 \left(\sum_{i=1}^{n}v_i^2\right),
\qquad \sum_i u_i=\sum_i v_i=0,
\]
with $\pi\sim {\rm Unif}(\Sf_n)$ and the bounded-score assumptions imply the extra factor $n^{-1}$ in \eqref{eq:app_eta_moment}; higher powers are handled by the same finite-population expansion and the fixedness of $q_j,q_k$.
Moreover,
\[
\E_0\{\eta_{sr}^{jk}\eta_{uv}^{\ell h}\}=0
\quad\hbox{whenever}\quad
|\{s,r\}\cap\{u,v\}|\le1.
\]
For disjoint pairs this follows from independence.  If the two pairs share one coordinate, conditioning on the shared rank sequence reduces the expectation to the product of two conditional means, each equal to zero by rank invariance.  Therefore, using \eqref{eq:app_linear_moment_bound}, \eqref{eq:app_eta_moment}, and
\[
\sum_{1\le s<r\le p}(p-r)^2\le Cp^4,
\qquad
\sigma_{q,n}^2=\npairs v_{V,q,n}\asymp p^2n^{-q},
\]
we obtain
\begin{align}
\E_0B_p(\bm a)
&=\sum_{j,k=1}^{K}a_ja_k\Gamma_{V,\calQ,n,jk}
\longrightarrow
\bm a^\top\mathbf{\Gamma}_{V,\calQ}\bm a,\label{eq:app_Bmean}\\
\Var_0\{B_p(\bm a)\}
&\le
C_{a,\calQ}
\sum_{j,k,\ell,h=1}^{K}
\frac{p^4 n^{-(q_j+q_k+q_\ell+q_h)/2-1}}
{p^4 n^{-(q_j+q_k+q_\ell+q_h)/2}}
\le C_{a,\calQ}n^{-1}.\label{eq:app_Bvar}
\end{align}
Thus $B_p(\bm a)\to_{P_0}\bm a^\top\mathbf{\Gamma}_{V,\calQ}\bm a$.  Furthermore, using \eqref{eq:app_linear_moment_bound} and the martingale Rosenthal inequality,
\begin{align}
\sum_{t=2}^{p}\E_0\{\xi_t(\bm{a})^4\}
&\le
C_{a,\calQ}
\sum_{t=2}^{p}
\left[
\frac{t^2\max_j v_{V,q_j,n}^{2}}{\min_j\sigma_{q_j,n}^{4}}
+
\frac{t\max_j n^{-2q_j}}{\min_j\sigma_{q_j,n}^{4}}
\right]                                             \notag\\
&\le C_{a,\calQ}
\left(
\frac{p^3\max_j n^{-2q_j}}{p^4\min_j n^{-2q_j}}
+
\frac{p^2\max_j n^{-2q_j}}{p^4\min_j n^{-2q_j}}
\right)
\le C_{a,\calQ}p^{-1}.                              \label{eq:app_lyap_rate}
\end{align}
Equations \eqref{eq:app_Bmean}--\eqref{eq:app_lyap_rate} imply the martingale CLT, because the conditional variance error is $O_{P_0}(n^{-1/2})$ and the Lyapunov term is $O(p^{-1})$. Hence $r_{V,\calQ}(n,p)\to0$, and Theorem \ref{thm:lq_clt} follows by the Cramer--Wold device.

\section{Proof of Theorem \ref{thm:block_independence}}\label{app:block}

All inequalities below are under $H_0$. Fix
\[
\calQ=\{q_1,\ldots,q_K\},\qquad \bm{x}=(x_1,\ldots,x_K)\in\mathbb R^K,
\qquad y\in\mathbb R.
\]
Let
\[
D_p(\bm{x})=\bigcap_{j=1}^{K}\{Z_{A,q_j}\le x_j\}.
\]
For $I=(s,t)\in\Lam$, define $E_I=E_I(y)$ by
\begin{equation*}
E_I=
\begin{cases}
\{|\wt A_{st}|>\sigma_{A,n}(4\log p-\log\log p+y)^{1/2}\},&A=V,W,\\[2mm]
\{\wt Q_{h,st}> {m\choose2}\lambda_1(n-1)^{-1}
[4\log p+(\mu_1-2)\log\log p-\Lambda/\lambda_1+y]\},&A=Q.
\end{cases}
\end{equation*}
Then
\begin{equation*}
\{M_A>y\}=\bigcup_{I\in\Lam}E_I.
\end{equation*}
For $r\ge1$ put
\begin{equation*}
H_A(p,r;y)=\sum_{I_1<\cdots<I_r\in\Lam}
\Pp_0(E_{I_1}\cdots E_{I_r}).
\end{equation*}

Under Assumption \ref{ass:V}, \ref{ass:W}, or \ref{ass:Q}, the moderate-deviation bound used in the rank-based maximum theory gives, for every fixed $y$,
\begin{equation}\label{eq:single_tail_bound_app}
\pi_A(p,y):=\max_{I\in\Lam}\Pp_0(E_I)
\le C_{A,y}p^{-2}.
\end{equation}
The condition needed for \eqref{eq:single_tail_bound_app} is
\begin{equation*}
\log p=o(n^{1/3})\quad(A=V,W),
\qquad
\log p=o(n^\theta)\quad(A=Q),
\end{equation*}
with $\theta$ as in \eqref{eq:theta_def}.
For $I_a=(s_a,t_a)$, split all ordered $r$-tuples $I_1<\cdots<I_r$ into
\begin{align*}
\Gamma_{p,1}(r)&=\{(I_1,\ldots,I_r): |\{s_a,t_a:1\le a\le r\}|=2r\},\\
\Gamma_{p,2}(r)&=\{(I_1,\ldots,I_r): s_1=\cdots=s_r\hbox{ or }t_1=\cdots=t_r\},\\
\Gamma_{p,3}(r)&=\{I_1<\cdots<I_r\}\setminus\{\Gamma_{p,1}(r)\cup\Gamma_{p,2}(r)\}.
\end{align*}
For $\Gamma_{p,1}(r)$ the exceedance events are independent, hence
\begin{equation*}
\sum_{\Gamma_{p,1}(r)}\Pp_0(E_{I_1}\cdots E_{I_r})
\le {\npairs\choose r}\{C_{A,y}p^{-2}\}^{r}
\le \frac{C_{A,y}^{r}}{2^r r!}.
\end{equation*}
Moreover,
\begin{equation*}
|\Gamma_{p,2}(r)|\le2p^{r+1},
\qquad
\sum_{\Gamma_{p,2}(r)}\Pp_0(E_{I_1}\cdots E_{I_r})
\le 2C_{A,y}^{r}p^{1-r}.
\end{equation*}
For $\Gamma_{p,3}(r)$, the conditional moderate-deviation argument used for overlapping pairs yields
\begin{equation*}
\sup_{(I_1,\ldots,I_r)\in\Gamma_{p,3}(r)}
\Pp_0(E_{I_1}\cdots E_{I_r})
\le C_{A,y,r}p^{-2r}.
\end{equation*}
Since
\begin{equation*}
|\Gamma_{p,3}(r)|
\le \sum_{\kappa=r+1}^{2r-1}{p\choose\kappa}\kappa^{2r}
\le (2r)^{2r}p^{2r-1},
\end{equation*}
we obtain the explicit bound
\begin{equation}\label{eq:H_rate_app}
H_A(p,r;y)
\le
\frac{C_{A,y}^{r}}{2^r r!}
+2C_{A,y}^{r}p^{1-r}
+C_{A,y,r}(2r)^{2r}p^{-1}.
\end{equation}
Consequently, for every fixed $r$,
\begin{equation*}
\limsup_{n,p\to\infty}H_A(p,r;y)
\le \frac{C_{A,y}^{r}}{2^r r!},
\end{equation*}
and
\begin{equation*}
\lim_{m\to\infty}\limsup_{n,p\to\infty}H_A(p,m;y)=0.
\end{equation*}

Fix $r\ge1$ and $I_1<\cdots<I_r$. Let
\[
\mathcal V_r=\{s_a,t_a: I_a=(s_a,t_a),\,1\le a\le r\},
\qquad |\mathcal V_r|\le2r,
\]
and
\begin{equation*}
\Lambda_{p,r}^*=
\{(u,v)\in\Lam: u\in\mathcal V_r\hbox{ or }v\in\mathcal V_r\}.
\end{equation*}
Then
\begin{equation*}
|\Lambda_{p,r}^*|\le2rp.
\end{equation*}
For $q\in\calQ$ set
\begin{align*}
R_{q}^{(r)}&=\sum_{J\in\Lambda_{p,r}^*}\{\wt A_J^{q}-\mu_{A,q,n}\},        \\
S_{q}^{(-r)}&=S_{A,q}-R_q^{(r)}.                                     
\end{align*}
Let
\[
\sigma_{q,n}^2=\npairs v_{A,q,n}.
\]
For every even $\tau\ge4$, the rank moment inequality for a fixed dependency degree gives
\begin{equation*}
\E_0|R_q^{(r)}|^\tau
\le C_{A,q,\tau}(2rp)^{\tau/2}v_{A,q,n}^{\tau/2}.
\end{equation*}
Hence, for every $\varepsilon\in(0,1)$,
\begin{align}
\Pi_{p,r,\tau}(\varepsilon)
&:=
\Pp_0\left(
\max_{1\le j\le K}\left|R_{q_j}^{(r)}/\sigma_{q_j,n}\right|>\varepsilon
\right)                                                        \notag\\
&\le
\sum_{j=1}^{K}
\varepsilon^{-\tau}\sigma_{q_j,n}^{-\tau}
\E_0|R_{q_j}^{(r)}|^\tau                                      \notag\\
&\le
C_{A,\calQ,\tau}\varepsilon^{-\tau}
\left(\frac{2rp}{\npairs}\right)^{\tau/2}                       \notag\\
&\le C_{A,\calQ,\tau}\varepsilon^{-\tau} r^{\tau/2}(p-1)^{-\tau/2}
\le C_{A,\calQ,\tau}\varepsilon^{-\tau} r^{2\tau}p^{-\tau/2}.     \label{eq:Pi_bound_app}
\end{align}
Define the shifted events
\begin{align*}
D_{p,r}^{+}(\bm{x},\varepsilon)
&=
\bigcap_{j=1}^{K}
\left\{S_{q_j}^{(-r)}/\sigma_{q_j,n}\le x_j+\varepsilon\right\},      \\
D_{p,r}^{-}(\bm{x},\varepsilon)
&=
\bigcap_{j=1}^{K}
\left\{S_{q_j}^{(-r)}/\sigma_{q_j,n}\le x_j-\varepsilon\right\}.      
\end{align*}
Let $E_{\mathbf I}=E_{I_1}\cdots E_{I_r}$. Since $D_{p,r}^{\pm}$ depends only on coordinates outside $\mathcal V_r$, while $E_{\mathbf I}$ depends only on coordinates in $\mathcal V_r$,
\begin{equation}\label{eq:indep_removed_app}
\Pp_0(D_{p,r}^{\pm}(\bm{x},\varepsilon)E_{\mathbf I})
=
\Pp_0(D_{p,r}^{\pm}(\bm{x},\varepsilon))\Pp_0(E_{\mathbf I}).
\end{equation}
Moreover, from \eqref{eq:Pi_bound_app},
\begin{align*}
\Pp_0\{D_p(\bm{x})E_{\mathbf I}\}
&\le
\Pp_0\{D_{p,r}^{+}(\bm{x},\varepsilon)E_{\mathbf I}\}
+
\Pi_{p,r,\tau}(\varepsilon),                                      \\
\Pp_0\{D_p(\bm{x})E_{\mathbf I}\}
&\ge
\Pp_0\{D_{p,r}^{-}(\bm{x},\varepsilon)E_{\mathbf I}\}
-
\Pi_{p,r,\tau}(\varepsilon).                                      
\end{align*}
The same removal bound also gives
\begin{align}
\Pp_0\{D_{p,r}^{+}(\bm{x},\varepsilon)\}
&\le
\Pp_0\{D_p(\bm{x}+2\varepsilon\bm{1}_K)\}+
\Pi_{p,r,\tau}(\varepsilon), \notag\\
\Pp_0\{D_{p,r}^{-}(\bm{x},\varepsilon)\}
&\ge
\Pp_0\{D_p(\bm{x}-2\varepsilon\bm{1}_K)\}-
\Pi_{p,r,\tau}(\varepsilon).                                      \label{eq:Dminus_compare}
\end{align}
Put
\begin{equation*}
\Delta_{p,\varepsilon}(\bm{x})
=
\max_{\bm{\eta}\in\{-2\varepsilon,0,2\varepsilon\}^{K}}
\left|
\Pp_0\{D_p(\bm{x}+\bm{\eta})\}-\Pp_0\{D_p(\bm{x})\}
\right|.
\end{equation*}
Equations \eqref{eq:indep_removed_app}--\eqref{eq:Dminus_compare} imply
\begin{equation}\label{eq:single_tuple_factor_rate}
\left|
\Pp_0\{D_p(\bm{x})E_{\mathbf I}\}
-
\Pp_0\{D_p(\bm{x})\}\Pp_0(E_{\mathbf I})
\right|
\le
\Delta_{p,\varepsilon}(\bm{x})\Pp_0(E_{\mathbf I})
+2\Pi_{p,r,\tau}(\varepsilon).
\end{equation}
Summing \eqref{eq:single_tuple_factor_rate} over all $I_1<\cdots<I_r$ gives
\begin{align}
\zeta_A(p,r;x,y)
&:=
\sum_{I_1<\cdots<I_r}
\left|
\Pp_0\{D_p(\bm{x})E_{I_1}\cdots E_{I_r}\}
-\Pp_0\{D_p(\bm{x})\}\Pp_0(E_{I_1}\cdots E_{I_r})
\right|                                                        \notag\\
&\le
\Delta_{p,\varepsilon}(\bm{x})H_A(p,r;y)
+2{\npairs\choose r}\Pi_{p,r,\tau}(\varepsilon)                  \notag\\
&\le
\Delta_{p,\varepsilon}(\bm{x})H_A(p,r;y)
+C_{A,\calQ,r,\tau}\varepsilon^{-\tau}p^{2r-\tau/2}.             \label{eq:zeta_rate_app}
\end{align}
By Theorem \ref{thm:lq_clt},
\begin{equation*}
\Delta_{p,\varepsilon}(\bm{x})
\le
2r_{A,\calQ}(n,p)
+
\sup_{\|\bm{u}-\bm{x}\|_\infty\le2\varepsilon}
|\Phi_{\mathbf{\Gamma}_{A,\calQ}}(\bm{u})-\Phi_{\mathbf{\Gamma}_{A,\calQ}}(\bm{x})|.
\end{equation*}
Since a non-degenerate Gaussian distribution has a bounded density on compact rectangles,
\begin{equation}\label{eq:Delta_eps_rate}
\limsup_{n,p\to\infty}\Delta_{p,\varepsilon}(\bm{x})
\le C_{\Gamma,K,x}\varepsilon.
\end{equation}
Choose $\tau=6r$ in \eqref{eq:zeta_rate_app}. Then
\begin{equation*}
\limsup_{n,p\to\infty}\zeta_A(p,r;x,y)
\le C_{\Gamma,K,x}\varepsilon\frac{C_{A,y}^{r}}{2^r r!}.
\end{equation*}
Letting $\varepsilon\downarrow0$ yields
\begin{equation*}
\lim_{n,p\to\infty}\zeta_A(p,r;x,y)=0
\qquad(r\hbox{ fixed}).
\end{equation*}
The explicit $p$-rate before the last limit is
\begin{equation*}
\zeta_A(p,r;x,y)
\le
\Delta_{p,\varepsilon}(\bm{x})
\left\{\frac{C_{A,y}^{r}}{2^r r!}+2C_{A,y}^{r}p^{1-r}+C_{A,y,r}(2r)^{2r}p^{-1}\right\}
+C_{A,\calQ,r}\varepsilon^{-6r}p^{-r}.
\end{equation*}

Let
\[
B_p(y)=\bigcup_{I\in\Lam}E_I=\{M_A>y\}.
\]
For $m\ge1$, Bonferroni's inequality gives
\begin{align}
\left|
\Pp_0\{D_p(\bm{x})B_p(y)\}
-
\sum_{r=1}^{m}(-1)^{r+1}
\sum_{I_1<\cdots<I_r}
\Pp_0\{D_p(\bm{x})E_{I_1}\cdots E_{I_r}\}
\right|
&\le H_A(p,m+1;y),                                      \label{eq:Bonf_D_app}\\
\left|
\Pp_0\{B_p(y)\}
-
\sum_{r=1}^{m}(-1)^{r+1}H_A(p,r;y)
\right|
&\le H_A(p,m+1;y).                                      \label{eq:Bonf_B_app}
\end{align}
Combining \eqref{eq:zeta_rate_app}, \eqref{eq:Bonf_D_app} and \eqref{eq:Bonf_B_app}, for any $m\ge1$,
\begin{align*}
&\left|
\Pp_0\{D_p(\bm{x})B_p(y)\}
-
\Pp_0\{D_p(\bm{x})\}\Pp_0\{B_p(y)\}
\right|                                                   \\
&\quad\le
\sum_{r=1}^{m}\zeta_A(p,r;x,y)
+2H_A(p,m+1;y)                                            \\
&\quad\le
\sum_{r=1}^{m}
\left[
\Delta_{p,\varepsilon}(\bm{x})H_A(p,r;y)
+C_{A,\calQ,r,\tau_r}\varepsilon^{-\tau_r}p^{2r-\tau_r/2}
\right]
+2H_A(p,m+1;y).                                
\end{align*}
Set $\tau_r=6r$. Using \eqref{eq:H_rate_app} and \eqref{eq:Delta_eps_rate},
\begin{align*}
\limsup_{n,p\to\infty}
&\left|
\Pp_0\{D_p(\bm{x})B_p(y)\}
-
\Pp_0\{D_p(\bm{x})\}\Pp_0\{B_p(y)\}
\right|                                               \\
&\le
C_{\Gamma,K,x}\varepsilon
\sum_{r=1}^{m}\frac{C_{A,y}^{r}}{2^r r!}
+
2\frac{C_{A,y}^{m+1}}{2^{m+1}(m+1)!}.       
\end{align*}
Now
\begin{equation*}
\lim_{m\to\infty}\lim_{\varepsilon\downarrow0}
\limsup_{n,p\to\infty}
\left|
\Pp_0\{D_p(\bm{x})B_p(y)\}
-
\Pp_0\{D_p(\bm{x})\}\Pp_0\{B_p(y)\}
\right|=0.
\end{equation*}
Thus
\begin{equation}\label{eq:block_ind_prob_app}
\Pp_0\{D_p(\bm{x}),M_A>y\}
=
\Pp_0\{D_p(\bm{x})\}\Pp_0\{M_A>y\}+o(1).
\end{equation}
Together with Theorem \ref{thm:lq_clt} and the rank-based maximum limits, \eqref{eq:block_ind_prob_app} proves Theorem \ref{thm:block_independence}.

\section{Proof of Theorem \ref{thm:cct}}\label{app:cct}

Let
\[
C_a=\tan\{\pi(1/2-P_{A,a})\},
\qquad
\mathcal C_{A,\calB}=\sum_{a\in\calB}w_aC_a.
\]
For $t>0$,
\begin{equation*}
\Pp\{C_0>t\}=\frac12-\frac1\pi\arctan(t)
=\frac{1}{\pi t}+O(t^{-3}),
\end{equation*}
where $C_0$ is a standard Cauchy random variable. Let
\[
t_\alpha=\tan\{\pi(1/2-\alpha)\}.
\]
Then
\begin{equation*}
t_\alpha=(\pi\alpha)^{-1}\{1+O(\alpha^2)\},
\qquad
\frac12-\frac1\pi\arctan(t_\alpha)=\alpha.
\end{equation*}
The Cauchy combination theorem of \citet{LiuXie2020} gives, for the fixed set $\calB$,
\begin{equation*}
\Pp_0\{\mathcal C_{A,\calB}>t\}
=
\frac{1}{\pi t}\{1+\delta_A(t,n,p)\},
\qquad
\lim_{t\to\infty}\limsup_{n,p\to\infty}|\delta_A(t,n,p)|=0.
\end{equation*}
Therefore, with $t=t_\alpha$,
\begin{align*}
\Pp_0(P_{A,\calB}\le\alpha)
&=
\Pp_0(\mathcal C_{A,\calB}>t_\alpha)                         \\
&=
\frac{1}{\pi t_\alpha}\{1+\delta_A(t_\alpha,n,p)\}             \\
&=
\alpha\{1+O(\alpha^2)+\delta_A(t_\alpha,n,p)
      +O(\alpha^2\delta_A(t_\alpha,n,p))\}.        
\end{align*}
Equation \eqref{eq:cct_level} follows.

If for some $a_0\in\calB$,
\begin{equation*}
P_{A,a_0}\xrightarrow{P_1}0,
\end{equation*}
then
\begin{equation*}
C_{a_0}
=\tan\{\pi(1/2-P_{A,a_0})\}
=\{\pi P_{A,a_0}\}^{-1}\{1+O(P_{A,a_0}^{2})\}
\xrightarrow{P_1}+\infty.
\end{equation*}
For the remaining finite number of indices $a\ne a_0$,
\begin{equation*}
\min_{a\ne a_0}P_{A,a}\ge \eta_n
\quad\Longrightarrow\quad
\max_{a\ne a_0}|C_a|\le C\eta_n^{-1},
\end{equation*}
and taking $\eta_n\downarrow0$ with
\[
\Pp_1\{P_{A,a_0}\le \eta_n^2\}\to1
\]
gives
\begin{equation*}
\mathcal C_{A,\calB}
\ge w_{a_0}\{\pi P_{A,a_0}\}^{-1}\{1+o_{P_1}(1)\}
-C\eta_n^{-1}
\xrightarrow{P_1}+\infty.
\end{equation*}
Thus
\begin{equation*}
P_{A,\calB}
=\frac12-\frac1\pi\arctan(\mathcal C_{A,\calB})
\xrightarrow{P_1}0.
\end{equation*}
This proves Theorem \ref{thm:cct}.

\section{Large-sample leading constants for quick implementation}\label{app:leading}

For the degenerate examples, the simulations use the fixed-$n$ Monte Carlo calibration in \eqref{eq:deg_mc_muvar}.  For larger samples, the following leading constants provide a fast analytic approximation and a useful diagnostic check.  They are not used for the finite-sample size tables in Section \ref{sec:simulation}.

For asymptotically normal coefficients with $\Var_0(T_{12})=\sigma_{T,n}^2$, the leading normal-moment formulas are
\[
\E(T_{12}^4)\sim3\sigma_{T,n}^4,
\qquad
\Var(T_{12}^4)\sim96\sigma_{T,n}^8,
\]
\[
\E(T_{12}^6)\sim15\sigma_{T,n}^6,
\qquad
\Var(T_{12}^6)\sim10170\sigma_{T,n}^{12}.
\]
For Spearman's $\rho$, $\sigma_{\rho,n}^2=(n-1)^{-1}$.  For Kendall's $\tau$, $\sigma_{\tau,n}^2=2(2n+5)/\{9n(n-1)\}$.

For the degenerate examples, let
\begin{equation*}
\calJ_a=\sum_{i,j\ge1}\frac{a}{\pi^4 i^2j^2}(\xi_{ij}^2-1),
\qquad
\xi_{ij}\iid N(0,1).
\end{equation*}
For $D$, $a=3$ and $B_D={5\choose2}=10$.  For $R$, $a=6$ and $B_R={6\choose2}=15$.  For $\tau^\ast$, $a=6$ and $B_{\tau^\ast}={4\choose2}=6$.  The leading constants are
\[
\mu_{T,q,n}^{\mathrm{lead}}=\left(\frac{B_T}{n}\right)^q E(\calJ_{a_T}^q),
\qquad
v_{T,q,n}^{\mathrm{lead}}=\left(\frac{B_T}{n}\right)^{2q}\Var(\calJ_{a_T}^q).
\]
The cumulants of $\calJ_a$ are
\begin{equation*}
\kappa_r(\calJ_a)=2^{r-1}(r-1)!\frac{a^r\zeta(2r)^2}{\pi^{4r}},
\qquad r\ge2.
\end{equation*}
With $M_r(a)=E(\calJ_a^r)$, the recursion
\begin{equation*}
M_0(a)=1,
\qquad M_1(a)=0,
\qquad
M_\ell(a)=\sum_{r=2}^{\ell}{\ell-1\choose r-1}\kappa_r(\calJ_a)M_{\ell-r}(a)
\end{equation*}
gives the following constants:
\begin{align*}
M_4(3)&=\frac{193}{3307500},\qquad
M_6(3)=\frac{4903025107}{894811943025000},\\
\Var(\calJ_3^4)&=\frac{37007208536234}{36365716622000390625},\\
\Var(\calJ_3^6)&=
\frac{354578595091740477776790799277}{2455001464593158529792371440312500000}.
\end{align*}
Since $\calJ_6=2\calJ_3$, $M_r(6)=2^rM_r(3)$ and $\Var(\calJ_6^r)=2^{2r}\Var(\calJ_3^r)$.

\begin{table}[htbp]
\centering
\caption{Leading constants for the three degenerate examples. These are large-sample analytic constants only; the finite-sample simulations use the Monte Carlo calibration in \eqref{eq:deg_mc_muvar}.}\label{tab:deg_leading}
\resizebox{\textwidth}{!}{%
\begin{tabular}{lcccc}
\toprule
$T$ & $\mu_{T,4,n}^{\mathrm{lead}}$ & $v_{T,4,n}^{\mathrm{lead}}$ & $\mu_{T,6,n}^{\mathrm{lead}}$ & $v_{T,6,n}^{\mathrm{lead}}$ \\
\midrule
$D$ & $0.583522298\,n^{-4}$ & $101.7640019\,n^{-8}$ & $5.479391670\,n^{-6}$ & $1.444311135\times10^{5}\,n^{-12}$ \\
$R$ & $47.26530612\,n^{-4}$ & $6.676736161\times10^{5}\,n^{-8}$ & $3994.476528\,n^{-6}$ & $7.675661537\times10^{10}\,n^{-12}$ \\
$\tau^\ast$ & $1.209991837\,n^{-4}$ & $437.5665811\,n^{-8}$ & $16.36137586\,n^{-6}$ & $1.287762316\times10^{6}\,n^{-12}$ \\
\bottomrule
\end{tabular}%
}
\end{table}

\begin{table}[htbp]
\centering
\scriptsize
\setlength{\tabcolsep}{4.2pt}
\renewcommand{\arraystretch}{0.88}
\caption{Empirical sizes of degenerate rank-based tests using high-order asymptotic centering and variance formulas.}
\label{tab:size-degenerate-highorder}
\begin{tabular}{llrrrrrr}
\toprule
$n$ & Marginal distribution
& $L_2$ & $L_4$ & $L_6$ & $L_\infty$
& $L_{2,\infty}$ & $L_{2,4,6,\infty}$ \\
\midrule
\multicolumn{8}{l}{\emph{Hoeffding's $D$}}\\
\midrule
100 & $N(0,1)$ & 8.7 & 26.7 & 9.2 & 8.5 & 9.4 & 17.5\\
100 & $t_3/\sqrt{3}$ & 7.3 & 25.1 & 8.8 & 7.5 & 8.0 & 17.3\\
100 & $(\chi^2_1-1)/\sqrt{2}$ & 8.0 & 26.2 & 8.8 & 7.7 & 8.5 & 17.2\\
\addlinespace[1pt]
200 & $N(0,1)$ & 6.7 & 13.5 & 5.3 & 6.4 & 7.3 & 9.5\\
200 & $t_3/\sqrt{3}$ & 5.4 & 11.5 & 3.8 & 4.9 & 5.6 & 7.8\\
200 & $(\chi^2_1-1)/\sqrt{2}$ & 7.4 & 14.6 & 6.5 & 6.6 & 8.1 & 11.8\\
\addlinespace[1pt]
400 & $N(0,1)$ & 6.2 & 8.7 & 5.1 & 6.5 & 6.5 & 6.9\\
400 & $t_3/\sqrt{3}$ & 5.9 & 8.2 & 4.6 & 6.4 & 5.9 & 7.1\\
400 & $(\chi^2_1-1)/\sqrt{2}$ & 7.3 & 9.5 & 4.1 & 4.6 & 6.9 & 8.0\\
\addlinespace[1pt]
800 & $N(0,1)$ & 5.7 & 6.9 & 3.2 & 5.1 & 5.9 & 5.9\\
800 & $t_3/\sqrt{3}$ & 4.6 & 7.4 & 4.9 & 6.5 & 5.5 & 7.1\\
800 & $(\chi^2_1-1)/\sqrt{2}$ & 6.1 & 7.4 & 4.0 & 5.5 & 6.7 & 6.5\\
\midrule
\multicolumn{8}{l}{\emph{Blum--Kiefer--Rosenblatt's $R$}}\\
\midrule
100 & $N(0,1)$ & 7.8 & 10.0 & 4.2 & 4.4 & 7.4 & 9.0\\
100 & $t_3/\sqrt{3}$ & 5.5 & 9.7 & 3.7 & 4.3 & 5.3 & 7.0\\
100 & $(\chi^2_1-1)/\sqrt{2}$ & 6.3 & 9.5 & 2.6 & 3.3 & 5.5 & 7.0\\
\addlinespace[1pt]
200 & $N(0,1)$ & 6.1 & 6.8 & 2.9 & 3.9 & 5.3 & 6.3\\
200 & $t_3/\sqrt{3}$ & 4.9 & 4.8 & 2.4 & 2.7 & 4.1 & 4.8\\
200 & $(\chi^2_1-1)/\sqrt{2}$ & 6.9 & 7.9 & 3.6 & 4.3 & 6.7 & 7.4\\
\addlinespace[1pt]
400 & $N(0,1)$ & 6.2 & 6.9 & 4.4 & 5.3 & 5.5 & 6.3\\
400 & $t_3/\sqrt{3}$ & 5.2 & 7.0 & 4.0 & 5.4 & 5.6 & 5.9\\
400 & $(\chi^2_1-1)/\sqrt{2}$ & 6.2 & 7.0 & 3.5 & 3.8 & 5.7 & 6.5\\
\addlinespace[1pt]
800 & $N(0,1)$ & 5.4 & 5.1 & 3.1 & 4.0 & 5.4 & 5.9\\
800 & $t_3/\sqrt{3}$ & 4.5 & 6.4 & 3.5 & 5.0 & 4.8 & 5.1\\
800 & $(\chi^2_1-1)/\sqrt{2}$ & 6.1 & 6.3 & 3.7 & 5.3 & 6.2 & 6.0\\
\midrule
\multicolumn{8}{l}{\emph{Bergsma--Dassios--Yanagimoto's $\tau^\ast$}}\\
\midrule
100 & $N(0,1)$ & 8.1 & 11.2 & 4.9 & 5.1 & 8.0 & 10.1\\
100 & $t_3/\sqrt{3}$ & 6.0 & 11.0 & 4.5 & 5.3 & 5.6 & 8.2\\
100 & $(\chi^2_1-1)/\sqrt{2}$ & 6.5 & 11.4 & 3.4 & 4.3 & 6.0 & 8.3\\
\addlinespace[1pt]
200 & $N(0,1)$ & 6.1 & 7.9 & 3.2 & 4.3 & 5.9 & 6.5\\
200 & $t_3/\sqrt{3}$ & 4.9 & 5.3 & 2.3 & 3.2 & 4.2 & 5.2\\
200 & $(\chi^2_1-1)/\sqrt{2}$ & 6.9 & 9.5 & 4.4 & 5.0 & 7.2 & 8.4\\
\addlinespace[1pt]
400 & $N(0,1)$ & 6.2 & 7.2 & 4.9 & 5.3 & 6.0 & 6.2\\
400 & $t_3/\sqrt{3}$ & 5.6 & 6.7 & 4.1 & 5.4 & 5.9 & 6.3\\
400 & $(\chi^2_1-1)/\sqrt{2}$ & 6.5 & 7.5 & 3.6 & 3.7 & 5.8 & 6.6\\
\addlinespace[1pt]
800 & $N(0,1)$ & 5.6 & 5.3 & 3.0 & 4.2 & 5.4 & 5.8\\
800 & $t_3/\sqrt{3}$ & 4.6 & 6.5 & 3.9 & 5.6 & 4.5 & 5.6\\
800 & $(\chi^2_1-1)/\sqrt{2}$ & 6.1 & 6.4 & 3.7 & 5.3 & 6.2 & 5.9\\
\bottomrule
\end{tabular}

\vspace{2mm}
\begin{minipage}{0.96\textwidth}
\footnotesize
\emph{Note.} The nominal level is $5\%$. Entries are empirical rejection frequencies multiplied by $100$. The simulation design is the same as in the main size experiment with $p=100$ and $1{,}000$ Monte Carlo replications; the only difference is that the centering and variance of the degenerate $L_q$ statistics are computed using the high-order asymptotic formulas in Table~\ref{tab:deg_leading} rather than the fixed-$n$ Monte Carlo calibration.
\end{minipage}
\end{table}

To assess the finite-sample accuracy of the high-order approximations reported in Table~\ref{tab:deg_leading}, we conduct an additional small-scale size experiment. The simulation design is identical to that of the main size study, including the data-generating mechanisms, nominal level, and number of Monte Carlo replications. The only difference is that, for the degenerate rank correlations, the centering and variance of the finite-\(L_q\) statistics are computed using the high-order asymptotic formulas in Table~\ref{tab:deg_leading} rather than the fixed-\(n\) Monte Carlo calibration. The resulting empirical sizes are reported in Table~\ref{tab:size-degenerate-highorder}.

The high-order asymptotic calibration improves as $n$ increases, but it remains noticeably liberal for some moderate-sample settings. The distortion is most visible for Hoeffding's $D$ with the $L_4$ statistic when $n=100$, where the empirical size is around $25$--$27\%$; the corresponding $L_{2,4,6,\infty}$ combination is also inflated because it inherits this liberal component. For $R$ and $\tau^\ast$, the high-order calibration is more stable, and by $n=400$ or $800$ most empirical sizes are close to the nominal $5\%$ level. These results support using direct fixed-$n$ Monte Carlo calibration for degenerate rank correlations at moderate sample sizes, while the high-order asymptotic formulas are more appropriate as a fast approximation for larger $n$.


\begin{thebibliography}{99}

\bibitem[Anderson(2003)]{Anderson2003}
Anderson, T. W. (2003).
\newblock \emph{An Introduction to Multivariate Statistical Analysis}, 3rd ed.
\newblock Wiley, New York.


\bibitem[Bai et al.(2009)]{BaiJiangYaoZheng2009}
Bai, Z., Jiang, D., Yao, J.-F. and Zheng, S. (2009).
\newblock Corrections to LRT on large-dimensional covariance matrix by RMT.
\newblock \emph{Annals of Statistics} \textbf{37} 3822--3840.

\bibitem[Bergsma and Dassios(2014)]{BergsmaDassios2014}
Bergsma, W. and Dassios, A. (2014).
\newblock A consistent test of independence based on a sign covariance related to Kendall's tau.
\newblock \emph{Bernoulli} \textbf{20} 1006--1028.

\bibitem[Blum, Kiefer and Rosenblatt(1961)]{BlumKieferRosenblatt1961}
Blum, J. R., Kiefer, J. and Rosenblatt, M. (1961).
\newblock Distribution free tests of independence based on the sample distribution function.
\newblock \emph{Annals of Mathematical Statistics} \textbf{32} 485--498.

\bibitem[Cai and Jiang(2011)]{CaiJiang2011}
Cai, T. T. and Jiang, T. (2011).
\newblock Limiting laws of coherence of random matrices with applications to testing covariance structure and construction of compressed sensing matrices.
\newblock \emph{Annals of Statistics} \textbf{39} 1496--1525.

\bibitem[Chatterjee(2021)]{Chatterjee2021}
Chatterjee, S. (2021).
\newblock A new coefficient of correlation.
\newblock \emph{Journal of the American Statistical Association} \textbf{116} 2009--2022.

\bibitem[Drton, Han and Shi(2020)]{DrtonHanShi2020}
Drton, M., Han, F. and Shi, H. (2020).
\newblock High-dimensional consistent independence testing with maxima of rank correlations.
\newblock \emph{Annals of Statistics} \textbf{48} 3206--3227.

\bibitem[Feng et al.(2022)]{FengJiangLiuXiong2022}
Feng, L., Jiang, T., Liu, B. and Xiong, W. (2022).
\newblock Max-sum tests for cross-sectional dependence of high-dimensional panel data.
\newblock \emph{Annals of Statistics} \textbf{50} 1124--1143.

\bibitem[Gretton et al.(2005)]{GrettonEtAl2005}
Gretton, A., Herbrich, R., Smola, A., Bousquet, O. and Sch\"olkopf, B. (2005).
\newblock Kernel methods for measuring independence.
\newblock \emph{Journal of Machine Learning Research} \textbf{6} 2075--2129.

\bibitem[Gretton et al.(2008)]{GrettonEtAl2008}
Gretton, A., Fukumizu, K., Teo, C. H., Song, L., Sch\"olkopf, B. and Smola, A. J. (2008).
\newblock A kernel statistical test of independence.
\newblock In \emph{Advances in Neural Information Processing Systems 20} 585--592.

\bibitem[Han, Chen and Liu(2017)]{HanChenLiu2017}
Han, F., Chen, S. and Liu, H. (2017).
\newblock Distribution-free tests of independence in high dimensions.
\newblock \emph{Biometrika} \textbf{104} 813--828.

\bibitem[He et al.(2021)]{HeEtAl2021}
He, Y., Xu, G., Wu, C. and Pan, W. (2021).
\newblock Asymptotically independent U-statistics in high-dimensional testing.
\newblock \emph{Annals of Statistics} \textbf{49} 154--181.

\bibitem[Heller, Heller and Gorfine(2013)]{HellerHellerGorfine2013}
Heller, R., Heller, Y. and Gorfine, M. (2013).
\newblock A consistent multivariate test of association based on ranks of distances.
\newblock \emph{Biometrika} \textbf{100} 503--510.

\bibitem[Hoeffding(1948b)]{Hoeffding1948b}
Hoeffding, W. (1948b).
\newblock A non-parametric test of independence.
\newblock \emph{Annals of Mathematical Statistics} \textbf{19} 546--557.

\bibitem[Jiang and Yang(2013)]{JiangYang2013}
Jiang, T. and Yang, F. (2013).
\newblock Central limit theorems for classical likelihood ratio tests for high-dimensional normal distributions.
\newblock \emph{Annals of Statistics} \textbf{41} 2029--2074.

\bibitem[Jiang(2004)]{Jiang2004}
Jiang, T. (2004).
\newblock The asymptotic distributions of the largest entries of sample correlation matrices.
\newblock \emph{Annals of Applied Probability} \textbf{14} 865--880.

\bibitem[Jiang, Jiang and Yang(2012)]{JiangJiangYang2012}
Jiang, D., Jiang, T. and Yang, F. (2012).
\newblock Likelihood ratio tests for covariance matrices of high-dimensional normal distributions.
\newblock \emph{Journal of Statistical Planning and Inference} \textbf{142} 2241--2256.

\bibitem[Kendall(1938)]{Kendall1938}
Kendall, M. G. (1938).
\newblock A new measure of rank correlation.
\newblock \emph{Biometrika} \textbf{30} 81--93.

\bibitem[Leung and Drton(2018)]{LeungDrton2018}
Leung, D. and Drton, M. (2018).
\newblock Testing independence in high dimensions with sums of rank correlations.
\newblock \emph{Annals of Statistics} \textbf{46} 280--307.

\bibitem[Li and Xue(2015)]{LiXue2015}
Li, D. and Xue, L. (2015).
\newblock Joint limiting laws for high-dimensional independence tests.
\newblock arXiv:1512.08819.

\bibitem[Liu and Xie(2020)]{LiuXie2020}
Liu, Y. and Xie, J. (2020).
\newblock Cauchy combination test: A powerful test with analytic p-value calculation under arbitrary dependency structures.
\newblock \emph{Journal of the American Statistical Association} \textbf{115} 393--402.

\bibitem[Liu, Lin and Shao(2008)]{LiuLinShao2008}
Liu, W.-D., Lin, Z. and Shao, Q.-M. (2008).
\newblock The asymptotic distribution and Berry--Esseen bound of a new test for independence in high dimension with an application to stochastic optimization.
\newblock \emph{Annals of Applied Probability} \textbf{18} 2337--2366.

\bibitem[Mao(2014)]{Mao2014}
Mao, G. (2014).
\newblock A new test of independence for high-dimensional data.
\newblock \emph{Statistics \& Probability Letters} \textbf{93} 14--18.

\bibitem[Mao(2017)]{Mao2017}
Mao, G. (2017).
\newblock Robust test for independence in high dimensions.
\newblock \emph{Communications in Statistics - Theory and Methods} \textbf{46} 10036--10050.

\bibitem[Mao(2018)]{Mao2018}
Mao, G. (2018).
\newblock Testing independence in high dimensions using Kendall's tau.
\newblock \emph{Computational Statistics \& Data Analysis} \textbf{117} 128--137.

\bibitem[Nagao(1973)]{Nagao1973}
Nagao, H. (1973).
\newblock On some test criteria for covariance matrix.
\newblock \emph{Annals of Statistics} \textbf{1} 700--709.

\bibitem[Pan et al.(2020)]{PanEtAl2020}
Pan, W., Wang, X., Zhang, H., Zhu, H. and Zhu, J. (2020).
\newblock Ball covariance: A generic measure of dependence in Banach space.
\newblock \emph{Journal of the American Statistical Association} \textbf{115} 307--317.

\bibitem[Pfister et al.(2018)]{PfisterEtAl2018}
Pfister, N., B\"uhlmann, P., Sch\"olkopf, B. and Peters, J. (2018).
\newblock Kernel-based tests for joint independence.
\newblock \emph{Journal of the Royal Statistical Society: Series B} \textbf{80} 5--31.

\bibitem[Roy(1957)]{Roy1954}
Roy, S. N. ({\color{red}1957}).
\newblock {\color{red}Some} aspects of multivariate analysis.
\newblock \emph{Wiley}, New York.

\bibitem[Schott(2005)]{Schott2005}
Schott, J. R. (2005).
\newblock Testing for complete independence in high dimensions.
\newblock \emph{Biometrika} \textbf{92} 951--956.

\bibitem[Shi, Xu and Du(2023)]{ShiXuDu2023}
Shi, X., Xu, M. and Du, J. (2023).
\newblock Max-sum test based on Spearman's footrule for high-dimensional independence tests.
\newblock \emph{Computational Statistics \& Data Analysis} \textbf{185} 107768.

\bibitem[Spearman(1904)]{Spearman1904}
Spearman, C. (1904).
\newblock The proof and measurement of association between two things.
\newblock \emph{American Journal of Psychology} \textbf{15} 72--101.

\bibitem[Sz\'ekely and Rizzo(2013)]{SzekelyRizzo2013}
Sz\'ekely, G. J. and Rizzo, M. L. (2013).
\newblock The distance correlation $t$-test of independence in high dimension.
\newblock \emph{Journal of Multivariate Analysis} \textbf{117} 193--213.

\bibitem[Sz\'ekely, Rizzo and Bakirov(2007)]{SzekelyRizzoBakirov2007}
Sz\'ekely, G. J., Rizzo, M. L. and Bakirov, N. K. (2007).
\newblock Measuring and testing dependence by correlation of distances.
\newblock \emph{Annals of Statistics} \textbf{35} 2769--2794.

\bibitem[Wang et al.(2024)]{WangLiuFengMa2024}
Wang, H., Liu, B., Feng, L. and Ma, Y. (2024).
\newblock Rank-based max-sum tests for mutual independence of high-dimensional random vectors.
\newblock \emph{Journal of Econometrics} \textbf{238} 105578.

\bibitem[Wang, Zou and Wang(2013)]{WangZouWang2013}
Wang, G., Zou, C. and Wang, Z. (2013).
\newblock A necessary test for complete independence in high dimensions using rank-correlations.
\newblock \emph{Journal of Multivariate Analysis} \textbf{121} 224--232.

\bibitem[Wu, Xu and Pan(2019)]{WuXuPan2019}
Wu, C., Xu, G. and Pan, W. (2019).
\newblock An adaptive test on high-dimensional parameters in generalized linear models.
\newblock \emph{Statistica Sinica} \textbf{29} 2163--2186.

\bibitem[Xia et al.(2025)]{XiaCaoDuDai2025}
Xia, L., Cao, R., Du, J. and Dai, J. (2025).
\newblock Consistent complete independence test in high dimensions based on Chatterjee correlation coefficient.
\newblock \emph{Statistical Papers} \textbf{66} 3.

\bibitem[Xia et al.(2026)]{XiaCaoDuLiu2025}
Xia, L., Cao, R., Du, J. and Liu, L. ({\color{red}2026}).
\newblock Rank-based combination independence tests for high-dimensional data.
\newblock \emph{Journal of Multivariate Analysis} \textbf{212} 105550.

\bibitem[Xu et al.(2016)]{XuLinWeiPan2016}
Xu, G., Lin, L., Wei, P. and Pan, W. (2016).
\newblock An adaptive two-sample test for high-dimensional means.
\newblock \emph{Biometrika} \textbf{103} 609--624.

\bibitem[Yao, Zhang and Shao(2018)]{YaoZhangShao2018}
Yao, S., Zhang, X. and Shao, X. (2018).
\newblock Testing mutual independence in high dimension via distance covariance.
\newblock \emph{Journal of the Royal Statistical Society: Series B} \textbf{80} 455--480.

\bibitem[Zhang, Wang and Shao(2025)]{ZhangWangShao2025}
Zhang, Y., Wang, R. and Shao, X. (2025).
\newblock Adaptive testing for high-dimensional data.
\newblock \emph{Journal of the American Statistical Association} \textbf{120} 1893--1905.


\end{thebibliography}
\end{document}